\documentclass[preprint,11pt,authoryear]{elsarticle}
\usepackage[utf8]{inputenc}
\usepackage{amsmath,amssymb,mathtools}
\journal{Applied Mathematics and Computation}
\usepackage{enumitem}
\usepackage{comment}
\usepackage{graphicx}
\usepackage[normalem]{ulem}
\usepackage{chemformula}
\usepackage{amsfonts}
\usepackage{textcomp}
\usepackage{bm}
\usepackage{appendix}
\usepackage{hyperref}

\usepackage{geometry}
\usepackage[poster=first]{animate}

\usepackage{lineno}

\usepackage{longtable}
\usepackage{multirow}
\usepackage{rotating}
\usepackage{framed, multido}

\begin{document}

\begin{frontmatter}
  
\title{Out-of-equlibrium inference of feeding rates through population data from generic consumer-resource stochastic dynamics}

\author{Jos\'e A. Capit\'an}
\address{Complex Systems Group, Department of Applied Mathematics, Universidad \unexpanded{Politécnica} de Madrid, Spain}
\ead{ja.capitan@upm.es (corresponding author)}
\author{David Alonso}
\address{Computational and Theoretical Ecology. CEAB-CSIC, Blanes, Catalonia,Spain}
\ead{dalonso@ceab.csic.es}

\date{\today}

\begin{abstract}
Statistical models are often structurally unidentifiable, because different sets of parameters can lead to equal model outcomes. To be useful for prediction and parameter inference from data, stochastic population models need to be identifiable, this meaning that model parameters can be uniquely inferred from a large number of model observations. In particular, precise estimation of feeding rates in consumer-resource dynamics is crucial, because consumer-resource processes are central in determining biomass transport across ecosystems. Model parameters are usually estimated at stationarity, because in that case model analyses are often easier. 
In this contribution we analyze the problem of parameter redundancy in a multi-resource consumer-resource model, showing that model indentifiability depends on whether the dynamics have reached stationarity or not. To be precise, we: (i) Calculate the steady-state and out-of-equilibrium probability distributions of predator's abundances analytically using generating functions, which allow us to unveil parameter redundancy and carry out proper maximum likelihood estimation. (ii) Conduct \emph{in silico} experiments by tracking the abundance of consumers that are either searching for or handling prey, data then used for maximum likelihood parameter estimation. (iii) Show that, when model observations are recorded out of equilibrium, feeding parameters are truly identifiable, whereas if sampling is done solely at stationarity, only ratios of rates can be inferred from data (i.e., parameters are redundant). We discuss the implications of our results when inferring parameters of general dynamical models. 
\end{abstract}


\end{frontmatter}



\section{Introduction}
\label{sec:intro}



The consumer-resource interaction is a key driver of the transport of biomass across ecological communities~\citep{belgrano2005aquatic,dunne2006network,getz2011biomass,perkins2022consistent,valdovinos2022bioenergetic}. In classical approaches, biomass flow is quantified by inter-dependent changes in species abundances, which are usually measured by the density of individuals of each species (or their biomass) in a given region~\citep{bersier2002quantitative}. This quantification has relied mainly on differential equations describing the spatial and temporal variability of species abundances. Consumer-resource interactions are included in these equations via mathematical functions called \emph{functional responses}, which encapsulate and simplify the complexities of feeding interactions, providing a coarse-grained representation of the underlying mechanisms~\citep{lafferty2015general,abrams2022food}. Functional responses are defined as the per capita, average feeding rate of consumers on their resources~\citep{berryman1992orgins}. Such functions have been derived from conceptual considerations about feeding dynamics (ranging from simple predator-prey equations to more complex food-web models~\citep{mcpeek2019mechanisms,abrams2022food}), but also from feeding experiments~\citep{Rosenbaum2018,barbier2021macro,abrams2022food}, where a consumer is observed feeding on a resource population. Holling's seminal papers~\citep{Holling1959a, Holling1959b} show how the behavior of individual consumers gives rise to an average feeding rate over the whole consumer population, which then can be used in population-level models. As shown in~\cite{palamara2021,palamara2022implicit}, Holling's type II functional response can be rigorously derived from processes at the individual level when mass action (individuals meet at random), chemostatic conditions (the total number of consumers and the total number of resources are kept constant) and stationarity are assumed. Functional responses have been generalized to multi-species systems~\citep{koen2007process,robertson2022accounting} and to include a variety of foraging mechanisms.


Elemental processes at the lower level of aggregation (i.e.,  individual level) in consumer-resource models are driven by feeding parameters. In the simplest scenario one can imagine, the feeding process involves two predator behavioral stages: the search for and the handling of preys. The search or hunting behavior of the available prey ends with a predator attack,  moment at which the predator starts handling the prey and stops searching for other preys. This predator-prey complex is usually irreversible (unless the prey escapes) and relaxes into a free predator again ready to search for more prey. This relaxation process takes some (handling) time, during which the individual predator kills, handles, and diggest the prey. Both attack and relaxation processes occur at certain rates that are not known a priori. These rates could be inferred from population abundances of the different types involved in the feeding process, given some hypotheses about how these abundances vary according to the two elemental processes. In particular, the distributions of the population numbers of each behavioral type (free and handling consumer numbers, in the example above) can be used to infer feeding parameters, provided that the statistical model is fitted to empirical abundance data. In this contribution, we focus on the problem of parameter inference using stochastic population models and \emph{in silico} (simulated) abundance data. 

Stochastic models are to be designed so that model parameters can be \emph{uniquely} determined from a large number of model realizations. In that case, the model is known as \emph{structurally identifiable}, because model parameters can be inferred without redundancy from data, regardless the quantity or quality of the data~\citep{bellman1970structural}. In particular, dynamic consumer-resource models should allow per-capita feeding rates (associated to elemental feeding processes) to be estimated unambiguously from experimental population abundance data. The problem of structural identifiability is general and of multidisciplinar character. For example, it has been thoroughly analyzed for deterministic dynamics in the context of biological modeling approaches (see~\cite{villaverde2016structural} and references therein), but also in social sciences~\citep{helbing2015saving,jusup2022social}. In this contribution, we focus on the problem of structural identifiability in stochastic dynamical models. Our aim is to unveil scenarios for feeding rate redundancy using a multi-resource, Holling type II feeding dynamics~\citep{Holling1959a,Holling1959b}. For that purpose, we consider a single predator population, in which individuals are able to search for and handle $S$ different resource species, leading to $S+1$ different behavioral types, namely: $S$ types associated to predators handling each resource species, and free predators that are not handling any resource. We assume mass action and chemostatic conditions to derive the joint probability distribution of the different behavioral types, at two regimes: in the transient dynamics (out of equilibrium) and in the long run (once the steady-state is reached). We show analytically that, at equilibrium, the model is not identifiable because only the ratio between relaxation and attack rates can be extracted from fitting model to data. In this sense, the model at equilibrium is \emph{set identifiable}, this meaning that parameters can be precisely determined only on a subset of the parameter space. We also show that true feeding rates cannot be properly estimated via a standard fitting procedure such as maximum likelihood (ML) parameter estimation using synthetic (model-simulated) data at stationarity. This \emph{in silico} experiment, which mimics an empirical setup to estimate feeding parameters that assumes the system has reached stationarity, only allows for precise estimation of quotients of rates at equilibrium.

However, if model observations are taken out of equilibrium at different sampling times, feeding parameters can be uniquely estimated via ML. In this case, the likelihood function associated to independent observations of the system's state exhibits a maximum at precise parameter values, making the model completely identifiable. This allows for proper parameter inference from data, providing a practical solution for parameter redundancy at equilibrium. In this line, to avoid potential ambiguities in model calibration from time series data, we stress that parameter inference should be done out of equilibrium. We also show that, for the multi-resource feeding dynamics studied, parameter redundancy arises if sampling is done for independent dynamical observations taken at the exact same time (either close to equilibrium or not). In actual experiments, this is not a shortcoming for inference. 

In order to carry out the analysis, we solve the stochastic model at equilibrium by computing the joint probability distribution of behavioral types, which turns out to be a multinomial distribution. In addition, we solve the master equation to obtain the temporal dependence of the joint probability distribution in the case of equal relaxation rates, which leads also to a multinomial with probability parameters depending on time. Thanks to these calculations, we can conduct parameter inference, both at equilibrium and out-of-equilibrium, using synthetic datasets based on model observations. 
The paper is organized as follows. Section~\ref{sec:gfun} summarizes the background on parameter redundancy and shows that the explicit computation of the generating function can be used to assess parameter redundancy. The analytical solution of the stochastic model is explained in detail in Section~\ref{sec:multi}. Section~\ref{sec:ident} is devoted to analyze several \emph{in silico} experiments designed to illustrate parameter identification failure, showing that feeding parameters can be only fully identifiable if the model is sampled out of equilibrium at different sampling times. At equilibrium, the model allows for the precise determination of dimensionless parameter ratios. Finally, we conclude discussing the implications and potential extensions of our work.

\section{Stochastic model identifiability via generating functions} 
\label{sec:gfun}

Precise inference of model parameters cannot be done for every statistical model: a model is defined as \emph{identifiable} if it is theoretically possible to infer true parameter values after retrieving an infinite number of observations from the model. Mathematically, this condition amounts to saying that different parameter sets lead to strictly different distributions of observable quantities pertaining to the sample space. More precisely, let $\mathcal{G}=(\mathcal{S},\mathcal{P})$ be a statistical model with sample space $\mathcal{S}$, onto which a vector of observable variables $\bm{x}$ takes values, and a set $\mathcal{P}$ of probability distributions $P_{\bm{\theta}}(\bm{x})$ whose parameter vector $\bm{\theta}$ is a member of the parameter space $\Theta$:
\begin{equation*}
\mathcal{P}=\{P_{\bm{\theta}}\vert \bm{\theta} \in \Theta\}.
\end{equation*}
Model $\mathcal{G}$ is \emph{identifiable} if the mapping $\bm{\theta} \mapsto P_{\bm{\theta}}$, which assigns a probability distribution to a set of parameters, is \emph{one-to-one}~\citep{lehmann2006theory}. As said, this means that distinct parameter sets should correspond to distinct probability distributions, i.e.: $P_{\bm{\theta}_1}(\bm{x})=P_{\bm{\theta}_2}(\bm{x}) \Leftrightarrow \bm{\theta}_1=\bm{\theta}_2$ almost surely for all $\bm{x}$. 

For an identifiable model, the map $\bm{\theta} \mapsto P_{\bm{\theta}}$ can be inverted to learn the true parameter set if the model can be observed an infinite number of times. This is true because of the strong law of large numbers, by which probabilities can be estimated by averaging frequencies over a large number of observations~\citep{loeve1977elementary}. Thus, with an infinite number of observations, finding the true probability distribution $P_{\bm{\theta}_0}$ is feasible. Because the map $\bm{\theta} \mapsto P_{\bm{\theta}}$ is invertible, the true parameter set $\bm{\theta}_0$ which generated the distribution $P_{\bm{\theta}_0}$ can be unambiguosly extracted from the distribution.

The easiest way to check for parameter redundancy is by finding an exhaustive summary~\citep{cole2020parameter}, which is a vector $\bm{\kappa}(\bm{\theta})$ of model parameter combinations that uniquely determine the statistical model $\mathcal{G}$, so that $\bm{\kappa}(\bm{\theta}_1)=\bm{\kappa}(\bm{\theta}_2) \Leftrightarrow P_{\bm{\theta}_1}(\bm{x})=P_{\bm{\theta}_2}(\bm{x})$ for all $\bm{\theta}_1$, $\bm{\theta}_2\in \Theta$. In general, any model will have many suitable exhaustive summaries. Knowledge of values of the exhaustive summary fully determines model outcomes. If the parameter vector $\bm{\theta}$ can be determined uniquely from an exhaustive summary $\bm{\kappa}(\bm{\theta})$, then the model lacks of parameter redundancy. See~\cite{castro2020testing} for more advanced ways of testing structural identifiability.

Stochastic model identifiability can be assessed if one can compute the distribution $P_{\bm{n}}(t)$ of observing state $\bm{n}$ at time $t$, which will depend on model parameters. For that purpose, the master (forward Kolmogorov) equation has to be solved~\citep{gillespie1991markov}, and from $P_{\bm{n}}(t)$ the exhaustive summary could be eventually extracted. Equivalently, one can obtain the multivariate $z$-transform of $P_{\bm{n}}(t)$, known as generating function, defined as
\begin{equation}\label{eq:Gdef}
G(\bm{z},t)=\sum_{n_1,\dots,n_S \ge 0} P_{\bm{n}}(t)\prod_{i=1}^S z_i^{n_i},
\end{equation}
where the vector $\bm{z}=(z_1,\dots,z_S)$ is conjugate to the state vector $\bm{n}=(n_1,\dots,n_S)$. This vector $\bm{z}$ can take values over $\mathbb{R}^S$. A series expansion will eventually lead to the probabilities $P_{\bm{n}}(t)$. Having $G(\bm{z},t)$ is equivalent to knowing $P_{\bm{n}}(t)$, because the generating function allows to compute every moment of the probability distribution, thus uniquely determining it. Therefore, a set of parameter combinations that uniquely determine the generating function can be used as an exhaustive summary to assess parameter redundancy.

We illustrate this idea using a well-known birth-death-immigration stochastic dynamics for a single population~\citep{kendall1949stochastic}. The elemental processes are defined by the following reaction scheme:
\begin{eqnarray}
\emptyset & \ch{->[ $\mu$ ]} & X, \label{BDI:immi}\\
X + \emptyset & \ch{->[ $\beta$  ]} & X + X,\label{BDI:birth}\\ 
X & \ch{->[ $\delta$  ]} & \emptyset.\label{BDI:death}
\end{eqnarray}
The first process stands for the arrival of a new individual $X$ by immigration at rate $\mu$, irrespectively of the number of individuals present in the system. The second one refers to reproduction at a per-capita rate $\beta$, this meaning that if there are $n$ individuals, they will reproduce with probability $\beta n$ per unit time. Finally, individual death occurs at a per-capita rate $\delta$. Since birth and death rates are linear on population numbers, the master equation can be solved analytically in terms of the generating function
\begin{equation*}
G(z,t)=\sum_{n\ge 0} P_{n}(t) z^n,
\end{equation*}
which results in~\citep{kendall1949stochastic}
\begin{equation*}
G(z,t)=\frac{\left(\frac{\beta}{\delta}-1\right)^{\frac{\mu}{\beta}}
\left[e^{(\beta-\delta)t}(1-z)+\frac{\beta}{\delta} z - 1\right]^{n_0}}
{\left[\frac{\beta}{\delta} e^{(\beta-\delta)t}(1-z)+\frac{\beta}{\delta} z -1\right]^{n_0+\frac{\mu}{\beta}}}
\end{equation*}
for an initial condition assuming $n(0)=n_0$ individuals at time $t=0$. We focus on the problem of estimating elemental process rates ($\beta$, $\delta$, and $\mu$) and assume that the initial condition $n_0$ is known in advance, so we do not need to infer it from data. The generating function ---and, henceforth, the probability $P_n(t)$--- is uniquely determined by specifying the parameter combinations $\mu/\beta$, $\beta-\delta$, and $\beta/\delta$. Thus, an exhaustive summary is $\bm{\kappa}(\bm{\theta})=(\mu/\beta,\beta-\delta,\beta/\delta)$. If vector $\bm{\kappa}(\bm{\theta})$ is known, then the parameter vector $\bm{\theta}=(\beta,\delta,\mu)$ can be uniquely determined and the model is identifiable. Therefore, the whole parameter set can be inferred from model observations via maximum likelihood parameter estimation (MLE) techniques, for example. 

On the other hand, let $\beta < \delta$ and assume that model observations are conducted at stationarity ($t\to\infty$). The generating function reduces to
\begin{equation}\label{eq:GBDI}
G(z,\infty)=\left(1-\frac{\beta}{\delta}\right)^{\frac{\mu}{\beta}}
\left(1-\frac{\beta}{\delta} z\right)^{-\frac{\mu}{\beta}}.
\end{equation}
Besides the trivial issue that $n_0$ cannot be inferred from steady-state observations in models with a single attractor like this one, there is a major issue in determining elemental process rates. According to~\eqref{eq:GBDI}, an exhaustive summary that fully determines steady-state model realizations is $\bm{\kappa}(\bm{\theta})=(\mu/\beta,\beta/\delta)$ for $\bm{\theta}=(\beta,\delta,\mu)$. Therefore, only the ratios $\mu/\beta$ and $\beta/\delta$ are truly identifiable, and the model remains parameter redundant (set identifiable) in this limit.

This simple example illustrates a problem that we expect to occur frequently, in stochastic population models, when it comes to estimating individual process rates. First, the master equation cannot be solved for general, multi-species birth-death processes when transition probability rates depend non-linearly on population numbers. Second, among the solvable models, it is much more frequent to find steady-state solutions of the master equation (thanks to detailed-balance conditions, for example) rather than the general temporal dependence of $P_{\bm{n}}(t)$. Therefore, having computed the steady-state solution $P_{\bm{n}}$ analytically, one could naively formulate a MLE method to estimate model rates by sampling the model at stationarity. As the example shows, this can lead to infinite possible estimates. Although model samples are generated at precise values of elemental process rates, they cannot be inferred from stationary distributions. The issue precisely stems by sampling the model after the steady state has been reached.

We conjecture this issue to be general (at least for models defined by elemental process rates with a polynomial dependence on population abundances), because at equilibrium there is no well-defined temporal scale, and the generating function could depend only on ratios of rates. Therefore, parameters should be estimated out of equilibrium. This has important implications for population models that are not analytically tractable, for which numerical log-likelihood profiles or simulation are to be applied. Moreover, if the MLE procedure is based on abundance data, one should be aware of whether data were collected at stationarity or not to be truly confident of ML estimates. In the rest of the paper, we demonstrate this pathology of MLE at equilibrium using a solvable multi-species consumer-resource model and show that the issue is resolved when data for MLE is collected out of equilibrium. In the discussion section, we highlight that natural extensions of this model would also be parameter redundant at equilibrium. 

\section{Multi-resource Holling Type II feeding dynamics} 
\label{sec:multi}

We start by defining a model for consumer-resource dynamics. Assume that an individual consumer $X_A$ can feed on a range of $S$ alternative resource species, $X_{R_i}$, with $i = 1, \dots, S$. We consider two processes: first, the attack of resource $X_{R_i}$ by the consumer $X_A$, which results in the formation of a compound $X_{AR_i}$ formed by the consumer handling the resource. Second, handling consumers do not search for prey until the compound $X_{AR_i}$ ``relaxes'' because the predator stops handling the prey, then becoming a consumer searching for prey. The stochastic dynamics can be summarized by $2\,S$ individual feeding reactions:
\begin{eqnarray}
X_A + X_{R_i} & \ch{->[ $\alpha_i$ ]} & \emptyset + X_{AR_i}, \label{int_AR_Multi_HII}\\
X_{AR_i}    & \ch{->[ $\nu_i$  ]}      & X_A. 
\label{death_AR_Multi_HII}
\end{eqnarray}
The first channel represents feeding interactions, by taking into account different resource preferences determined by distinct attack rates ($\alpha_i)$ for the resources. The second reaction models consumer-resource compound relaxation, which occurs at rates $\nu_i$ (depending on resource type). Mean handling times, given by $\nu_i^{-1}$, are also characteristic for individual consumers feeding on each resource type. Our aim in this section is to solve analytically the stochastic model, by obtaining the joint probability distribution of the number of searching predators, as well as the number of consumers handling on each resource, both out-of-equilibrium and at stationarity.

The total consumer population, $n_A^0$, is partitioned into $S+1$ classes, 
\begin{equation}\label{eq:const}
n_A^0 = n_A + \sum_{i=1}^S n_i,\qquad i=1,\dots,S,
\end{equation}
where $n_A$ refers to the number of (free) consumers searching for resources, and $n_i$ is the number of predators handling on resource $i$. In this section, we first write the reaction scheme (\ref{int_AR_Multi_HII})--(\ref{death_AR_Multi_HII}) as transition probability rates between discrete states, then solve the master equation at stationarity, and finally obtain the temporal dependence of the probability distribution in the particular case in which these two conditions hold: (i) relaxation rates are equal, $\nu_1=\dots=\nu_S$, and (ii) initially there are only $n_A=n_A^0$ free consumers. Either for finite times or as $t\to \infty$, we obtain multinomial probability distributions.

The master equation determines the temporal evolution of the probability of finding a given discrete configuration of consumer types, $(n_A, n_1, \dots, n_S)$, at a given time $t$. As in~\cite{palamara2021}, we assume chemostatic conditions, so the overall number of consumers, $n_A^0$, is kept constant. Therefore, the number of predators $n_A$ is immediately determined by the number of consumers handling on each resource, $\{n_i\}$, via Eq.~\eqref{eq:const}. Accordingly, we define the state vector of the stochastic process as $\bm{n}=(n_1,\dots,n_S)$, considering only variations in handling predator numbers. Given a configuration vector $\bm{n}$, the number of free predators is determined as $n_A=n_A^0-\sum_{i=1}^S n_i$. 

Let $\bm{e}_i=(0,\dots,1,\dots,0)$, $i=1,\dots,S$, be the vectors of the canonical basis of $\mathbb{R}^S$. Transition probability rates are defined as
\begin{equation*}
T(\bm{n}+\bm{e}_i|\bm{n}) = \alpha_i \frac{n_A n_{R_i}}{N},\qquad
T(\bm{n}-\bm{e}_i|\bm{n}) = \nu_i n_i, 
\end{equation*}
for $i=1,\dots,S$, each one corresponding to chemical reactions \eqref{int_AR_Multi_HII} and \eqref{death_AR_Multi_HII}, respectively. They account for the conditional probability for transition $\bm{n}\to \bm{n}+\bm{e}_i$, i.e, the event of a consumer that starts handling on resource $i$, as well as the conditional probability for transition $\bm{n}\to \bm{n}-\bm{e}_i$, which represents the relaxation of the compound formed by the consumer and the $i$-th resource. For handling predators formation, we assume also chemostatic conditions for resources levels, meaning that resource densities $x_{R_i}:=n_{R_i}/N$ are kept constant ---here $N$ stands for the number of sites of the system (system volume). In addition, a law of mass action is assumed because the probability of consumer-resource encounter has to be proportional to resource availability. The effective per capita attack rate of a free consumer at a given density level of the $i$-th resource type is a resource-specific constant parameter defined by $\theta_i:=\alpha_i x_{R_i}$. The second transition probability represents $AR_i$ compounds relaxation, leading to free predators after a mean handling time $\nu_i^{-1}$. This probability is proportional to the number of handling predators $n_i$ if one of the $AR_i$ compounds stops handling the resource. 

Using Eq.~\eqref{eq:const}, transition probability rates become
\begin{equation}\label{eq:rates}
T(\bm{n}+\bm{e}_i|\bm{n}) = \theta_i \bigg(n_A^0-\sum_{j=1}^S n_j\bigg),\qquad
T(\bm{n}-\bm{e}_i|\bm{n}) = \nu_i n_i, 
\end{equation}
where $n_A^0$ and $\nu_i$, $\theta_i$ for $i=1,\dots,S$, are constant parameters. Let $P_{\bm{n}}(t)$ be the probability of observing a state vector $\bm{n}$ at time $t$. The master equation expresses the balance of probabilities as an infinite system of coupled differential equations,
\begin{equation}\label{eq:master}
\frac{d P_{\bm{n}}}{dt} = \sum_{i=1}^S\left\{
T(\bm{n}|\bm{n}+\bm{e_i})P_{\bm{n}+\bm{e_i}}+
T(\bm{n}|\bm{n}-\bm{e_i})P_{\bm{n}-\bm{e_i}}-
\left[T(\bm{n}+\bm{e_i}|\bm{n})+T(\bm{n}-\bm{e_i}|\bm{n})\right]P_{\bm{n}}\right\}.
\end{equation}
The probability balance can be interpreted as follows. The probability of visiting state $\bm{n}$ at time $t$ increases at rate $T(\bm{n}|\bm{n}-\bm{e_i})=\theta_i \left(n_A^0-(n_i-1)-\sum_{j\ne i}^S n_j\right)$ for the transition $\bm{n}-\bm{e}_i\to \bm{n}$, where a free consumer starts handling a resource of type $i$. The probability $P_{\bm{n}}(t)$ also increases through the transition $\bm{n}+\bm{e}_i\to \bm{n}$ at rate $T(\bm{n}|\bm{n}+\bm{e}_i) = \nu_i (n_i+1)$, when a handling consumer leaves the resource and starts searching for other resources. The negative terms in the right-hand side of~\eqref{eq:master} simply balance the transitions from $\bm{n}$ to any other state, which force the probability $P_{\bm{n}}(t)$ to decrease.

Our aim is to accurately estimate feeding rates $\{\alpha_1,\dots,\alpha_S,\nu_1,\dots,\nu_S\}$. We first compute the steady-state distribution for arbitrary parameter values, and show that the distribution of feeding types is a multinomial $\mathcal{M}(n_A^0,\{p_A,p_1,\dots,p_S\})$ with index $n_A^0$ and probability parameters $\{p_A,p_1,\dots,p_S\}$. Labeling $p_0:=p_A$, the multinomial density function is defined as
\begin{equation}\label{eq:multidef}
P_{\bm{n}}=
\begin{cases}
\frac{n_A^0!}{\prod_{k=0}^S n_k!} \prod_{k=0}^S p_k^{n_k}, & 
\text{when }n_A+\sum_{k=1}^S n_k = n_A^0,\\
0, & \text{otherwise}.
\end{cases}
\end{equation}
Here, vector $\bm{n}=(n_A,n_1,\dots,n_S)$ gathers the number of searching predators, $n_A$, together with the numbers $n_i$ of predators handling on each resource ($i=1,\dots,S$). Probability parameters are
\begin{equation}\label{eq:pi}
p_i = \frac{\theta_i}{\nu_i}\frac{1}{1+\sum_{k=1}^S \frac{\theta_k}{\nu_k}}
\end{equation}
for $i=1,\dots,S$, and 
\begin{equation}\label{eq:pA}
p_A = p_0 = \frac{1}{1+\sum_{k=1}^S \frac{\theta_k}{\nu_k}}=1-\sum_{k=1}^S p_k
\end{equation}
is the probability associated to the number of free predators, $n_A=n_A^0-\sum_{i=1}^S n_i$. To carry out the derivation, we use a detailed-balance condition satisfied by transition probabilities. The details can be found in~\ref{sec:appA00}. 

Second, using the generating function formalism, we find an analytical solution for out-of-equilibrium probabilities $P_{\bm{n}}(t)$ for the particular case of equal relaxation rates, $\nu_1=\dots=\nu_S$, and for an initial state formed only by searching predators. In the second case, we also show that the distribution is multinomial. From the form of the distributions (and the generating function), it is apparent that the model is perfectly identifiable out of equilibrium.

In the general case of different relaxation rates, we cannot analytically solve the master equation. However, it is a linear system of differential equations, so it could be easily integrated numerically. Besides numerical integration, there are other exact methods ---based on matrix exponentiation~\citep{keeling2009efficient} or Laplace transforms~\citep{etienne2007neutral}--- or approximate ---dimensionality reduction~\citep{keeling2009efficient}--- that allow the computation of probabilities over time. Although the parameter estimation analysis presented below would be numerically feasible in the case of distinct relaxation rates, we do not discuss it in this contribution.

\subsection{Out-of-equilibrium probability distribution}

Calculating the out-of-equilibrium probability distribution involves integrating the master equation~\eqref{eq:master} as a function of time. Usually this equation has to be integrated numerically. In this case, however, an analytical solution is feasible because transition probability rates depend linearly on population numbers~\citep{kendall1949stochastic,nisbet2004modelling}. We solve the master equation by introducing the generating function defined in Eq.~\eqref{eq:Gdef}. Transforming $P_{\bm{n}}(t)$ into the generating function $G(\bm{z},t)$ converts the infinite system of coupled differential equations into a partial differential equation.

We start by obtaining the PDE satisfied by the generating function $G(\bm{z},t)$ defined in Eq.~\eqref{eq:Gdef}. 
By multiplying the master equation, Eq.~\eqref{eq:master}, by the product $z_1^{n_1}\cdots z_S^{n_S}$, and summing over the population variables $n_i$ ($i=1,\dots,S$) over non-negative integers, we obtain the following PDE for the generating function $G$,
\begin{equation*}
\frac{\partial G}{\partial t} = \sum_{i=1}^S (1-z_i)
\biggl(\nu_i\frac{\partial G}{\partial z_i}
+\theta_i \sum_{j=1}^S z_j\frac{\partial G}{\partial z_j}
-\theta_i n_A^0 G\biggr).
\end{equation*}
Equivalently, it can be rewritten as
\begin{equation}\label{eq:PDE}
\frac{\partial G}{\partial t} = -n_A^0 m(\bm{z}) G +\sum_{i=1}^S 
[(1-z_i)\nu_i+z_i m(\bm{z})]\frac{\partial G}{\partial z_i},
\end{equation}
where we have introduced the notation
\begin{equation}\label{eq:m}
m(\bm{z}):=\sum_{j=1}^S \theta_j(1-z_j).
\end{equation}
We leave the details of the derivation of Eq.~\eqref{eq:PDE} from~\eqref{eq:master} to~\ref{sec:appA0}. 

We assume that the initial condition for the probability distribution is the following, without loss of generality:
\begin{equation*}
P_{\bm{n}}(0)=\prod_{i=1}^S \delta_{n_i,\hat{n}_i},
\end{equation*}
i.e., initially the state of the system is defined by the population vector $\bm{\hat{n}}=(\hat{n}_1,\dots,\hat{n}_S)$, where $\hat{n}_i$ stands for the initial number of predators handling resource $i$, and there are $\hat{n}_A=n_A^0-\sum_{j=1}^S \hat{n}_j$ free predators ---we assume that $n_A^0-\sum_{j=1}^S \hat{n}_j\ge 0$ holds. In the initial condition above, $\delta_{n,m}$ stands for the Kronecker symbol, i.e., $\delta_{n,m}=1$ if $n=m$ and $0$ otherwise. In terms of $G(\bm{z},t)$, the initial condition reduces to
\begin{equation}\label{eq:Gini}
G(\bm{z},0)=\prod_{i=1}^S z_i^{\hat{n}_i}.
\end{equation}
It is important to remark that Eq.~\eqref{eq:PDE} immediately implies that the probability distribution $P_{\bm{n}}(t)$ is normalized for all $t$: indeed, if $z_i=1$ for all $i$, then if follows from~\eqref{eq:PDE} that $\frac{\partial}{\partial t} G(\bm{1},t)=0$, i.e., $G(\bm{1},t)$ must be a constant ---here we use the notation $\bm{1}=(1,\dots,1)$ for a vector of ones. Thus, by definition of $G(\bm{z},t)$, 
\begin{equation*}
G(\bm{1},t)=\sum_{n_1,\dots,n_S\ge 0} P_{\bm{n}}(t) = 1,
\end{equation*}
which follows from the normalization of the probability distribution. Therefore, the condition $G(\bm{1},t)=1$ is satisfied by the PDE and automatically implies the normalization of the distribution.

Now we solve the first-order PDE~\eqref{eq:PDE} using the method of characteristics curves. An equivalent way of writing the initial PDE in a very compact form is as the following dot product, $\bm{v}\cdot\bm{\nabla}H=0$, where the solution $g=G(\bm{z},t)$ is viewed in the $(S+2)$-dimensional space $(\bm{z},t,g)$ as the manifold $H(\bm{z},t,g)=G(\bm{z},t)-g=0$, and the vector field $\bm{v}$ is defined as
\begin{equation*}
\bm{v}=(\bm{w},-1,n_A^0 m(\bm{z}) G(\bm{z},t)).
\end{equation*}
Here the row vector $\bm{w}=(w_i)$ has $S$ components defined as $w_i=\nu_i(1-z_i)+  m(\bm{z})z_i$, for $i=1,\dots,S$. Finding the integral surface $H$ amounts to solving the PDE.

Since the gradient of $H$ is normal to the integral surface at each and every point,
$(\bm{z},t,g)$, then any curve on the surface should have a tangential direction, defined by
the vector $\bm{v}$, that should be perpendicular to $\bm{\nabla}H$ at every point. Now, we write an arbitrary curve over the manifold $H=0$ in parametric form as $(\bm{z}(s),t(s),g(s))$, and the tangent vector $(\bm{z}'(s),t'(s),g'(s))$ has to be parallel to vector $\bm{v}$: this immediately yields the following non-linear system of $S+1$ coupled ODEs,
\begin{equation}\label{eq:char}
\begin{aligned}
\frac{dz_i}{dt} &= -\nu_i(1-z_i)-m(\bm{z}) z_i,\quad\text{for }i=1,\dots,S,\\
\frac{dG}{dt} &= -n_A^0 m(\bm{z}) G.
\end{aligned}
\end{equation}
where we have chosen $s=-t$ and used that $\frac{dg}{dt}=\frac{dG}{dt}$ because $g$ equals $G(\bm{z},t)$ over any characteristic curve within the manifold $H=0$. 

This system could be easily solved if we knew the implicit time dependence of function $m(t):=m(\bm{z}(t))$, see Eq.~\eqref{eq:m}, because $z_i$ are indeed functions of time. Thus, the ODE system can be expressed by adding an extra equation for $m(t)$ as an additional variable, which is obtained by multiplying the $i$-th equation in~\eqref{eq:char} by $\theta_i$ and summing over all the equations for $z_i$. This process yields
\begin{equation}\label{eq:edom}
\frac{dm}{dt}=-m^2+Qm+\sum_{j=1}^S \nu_j\theta_j(1-z_j),
\end{equation}
for $Q:=\sum_{j=1}^S \theta_j$. The coupled ODE system to solve for the time dependence of the $S+2$ unknowns $(\bm{z},m,G)$ is
\begin{equation}\label{eq:char1}
\begin{aligned}
\frac{dz_i}{dt} &= -\nu_i(1-z_i)-m z_i,\quad\text{for }i=1,\dots,S,\\
\frac{dm}{dt} &= -m^2 + Qm + \sum_{j=1}^S \nu_j\theta_j(1-z_j),\\
\frac{dG}{dt} &= -n_A^0 m\,G.
\end{aligned}
\end{equation}

In the particular case in which all the relaxation rates are equal, $\nu_1=\dots=\nu_S=:\nu$, the system decouples because the second equation is closed for $m$, according to Eq.~\eqref{eq:m}, so we can find the time dependence of function $m(t):=m(\bm{z}(t))$. The condition of equal relaxation rates amounts to assuming that the predator spends the same average time $\nu^{-1}$ handling any type of resource. Therefore, the equation for $m(t)$ reduces to a logistic growth equation if the assumption of equal handling times (equal relaxation rates) holds:
\begin{equation}\label{eq:EDOm}
\frac{dm}{dt}=-m^2+(\nu+Q)m.
\end{equation}
From now on in this section we will assume that relaxation rates $\nu_j$ are equal, in order to find explicit, closed-form solutions for the PDE satisfied by the generating function.

Integrating Eq.~\eqref{eq:EDOm} yields 
\begin{equation}\label{eq:mt}
m(t)=\frac{\nu+Q}{1+K e^{-(\nu+Q)t}},
\end{equation}
for $K$ an arbitrary constant to be calculated as function of the initial conditions, $\hat{z}_j:=z_j(0)$. Then solving for $K$ in Eq.~\eqref{eq:mt} at $t=0$ gives 
\begin{equation}\label{eq:K}
K=\frac{\nu+R}{Q-R}=\frac{\nu+\sum_{j=1}^S \theta_j\hat{z}_j}{\sum_{j=1}^S \theta_j(1-\hat{z}_j)},
\end{equation}
where we have defined $R:=\sum_{j=1}^S \theta_j\hat{z}_j$. Observe that the initial condition vector $\bm{\hat{z}}$ depends on the vector of $z$-variables, $\bm{z}$, via the explicit solution of the differential equations for $z_i$ in Eq.~\eqref{eq:char}. Because the solution of the system of ODEs satisfied by the characteristic curves will depend on the vector $\bm{\hat{z}}$ of initial conditions, we have to substitute $\hat{z}_j$ as function of $\bm{z}$ at the end of the calculation in order to get the desired expression for the generating function $G(\bm{z},t)$.

Therefore, in the particular case of equal handling times, the system of ODEs~\eqref{eq:char1} turns out to be uncoupled. Upon substitution of Eq.~\eqref{eq:mt} into~\eqref{eq:char1}, it is immediate to solve the equation for $G$, which gives
\begin{equation*}
G(\bm{\hat{z}},t)=G_0(K+e^{(\nu+Q)t})^{-n_A^0},
\end{equation*}
where $K$ and $G_0$ depend on the initial conditions $\bm{\hat{z}}$. Imposing the initial condition $G(\bm{\hat{z}},0)$ given by Eq.~\eqref{eq:Gini}, we deduce that $G_0=(1+K)^{n_A^0}\prod_{j=1}^S \hat{z}_j^{\hat{n}_j}$, and
\begin{equation}\label{eq:Gs}
G(\bm{\hat{z}},t)=\left(\frac{1+K}{K+e^{(\nu+Q)t}}\right)^{n_A^0}
\prod_{j=1}^S \hat{z}_j^{\hat{n}_j}.
\end{equation}
Recall that the desired dependence on $\bm{z}$ is obtained once we make the substitution $\hat{z}_j=\hat{z}_j(\bm{z},t)$. So it only remains to find the dependence of the initial conditions $\hat{z}_j$ in terms of $\bm{z}$. 

From Eq.~\eqref{eq:mt} we observe that the ODEs for $z_i(t)$ are equivalent, so we have to solve just one of the equations, for example, the one for $z_j$:
\begin{equation*}
\frac{d z_j}{dt}=-(1-z_j)\nu-\frac{(\nu+Q)z_j}{1+Ke^{-(\nu+Q)t}},
\end{equation*}
subject to the initial condition $z_j(0)=\hat{z}_j$. Then we immediately find that
\begin{equation}\label{eq:zj}
z_j(t)=\frac{Q K-\nu e^{(\nu+Q)t}+e^{\nu t}
[\nu+Q\hat{z}_j-Q K(1-\hat{z}_j)]}{Q(K+e^{(\nu+Q)t})},
\end{equation}
Multiplying this expression by $\theta_j$ and summing over $j=1,\dots,S$, we get
\begin{equation*}
Q-m(\bm{z})=\frac{KQ-\nu e^{(\nu+Q)t}+e^{\nu t}[(1+K)R+\nu-KQ]}{K+e^{(\nu+Q)t}},
\end{equation*}
because $\sum_{j=1}^S \theta_j z_j=Q-m(\bf{z})$ by~\eqref{eq:m}. Given that $\hat{z}_j$ are functions of $\bm{z}$, $K$ becomes itself a function of $\bm{z}$, which can be determined as follows. From Eq.~\eqref{eq:K} we can solve for $R=\sum_{j=1}^S \theta_j\hat{z}_j$ in terms of $K$ as $R=(KQ-\nu)/(1+K)$ which, upon substitution into the last expression, yields 
\begin{equation*}
Q-m(\bm{z})=\frac{KQ-\nu e^{(\nu+Q)t}}{K+e^{(\nu+Q)t}}.
\end{equation*}
In this equality, any dependence on the initial condition $\bm{\hat{z}}$ has been removed. We can solve for $K$ in terms of $m$, giving
\begin{equation}\label{eq:Kz}
K=K(\bm{z},t)=\left(\frac{\nu+Q}{m(\bm{z})}-1\right)e^{(\nu+Q)t}.
\end{equation}
Finally, an explicit expression of initial conditions $\hat{z}_j$ as function of $\bm{z}$ is determined by substituting $K(\bm{z},t)$, given by Eq.~\eqref{eq:Kz}, into~\eqref{eq:zj}, which results in an equation from which we can eliminate $\hat{z}_j$ in terms of $\bm{z}$. This manipulation gives
\begin{equation}\label{eq:hatz}
\hat{z}_j=1-\left(1+\frac{\nu}{Q}\right)
\frac{m(\bm{z})+[(1-z_j)Q-m(\bm{z})]e^{Qt}}{m(\bm{z})+[\nu+Q-m(\bm{z})]e^{(\nu+Q)t}}.
\end{equation}

Substituting Eqs.~\eqref{eq:Kz} and \eqref{eq:hatz} into Eq.~\eqref{eq:Gs} yields the final expression for the generating function, expressed in closed form as function of $\bm{z}$:
\begin{equation}\label{eq:Gtgen}
G(\bm{z},t)=\left[1+\frac{m({\bm{z}})(e^{-(\nu+Q)t}-1)}{\nu+Q}\right]^{n_A^0}
\prod_{j=1}^S\left[1-\left(1+\frac{\nu}{Q}\right)
\frac{m(\bm{z})+[Q (1-z_j)-m(\bm{z})]e^{Qt}}{m(\bm{z})+[\nu+Q-m(\bm{z})]e^{(\nu+Q)t}}\right]^{\hat{n}_j}.
\end{equation}
Recall the definitions $m(\bm{z}):=\sum_{j=1}^S \theta_j (1-z_j)$ and $Q:=\sum_{j=1}^S \theta_j$, and that the condition $n_A^0-\sum_{j=0}^S \hat{n}_j\ge 0$ must hold for the last expression to make sense. This function satisfies the initial condition~\eqref{eq:Gini} as well as the normalization condition $G(\bm{1},t)=1$. The generating function has to be expanded in power series of $z_i$ in order to obtain the explicit expression of the probability distribution $P_{\bm{n}}(t)$. Such task is hard in general. As shown by the derivation itself, it is very difficult to find the out-of-equilibrium distribution from the master equation, in contrast with the solution at stationarity: for the former, we cannot find it unless assuming equal handling rates, and we cannot carry out the series expansion to obtain the probability distribution $P_{\bm{n}}(t)$ for arbitrary initial conditions.

However, the series expansion can be carried out for the particular case $\hat{n}_1=\dots=\hat{n_S}=0$, which means that, initially, the population is formed solely by $n_A=n_A^0$ free predators and no individual consumer is initially handling any prey. In this case, Eq.~\eqref{eq:Gtgen} reduces to
\begin{equation}\label{eq:Gt}
G(\bm{z},t)=\biggl(p_0(t)+\sum_{j=1}^S p_j(t) z_j\biggr)^{n_A^0},
\end{equation}
with
\begin{equation}\label{eq:pj}
p_0(t)=\frac{1+e^{-\left(\nu+\sum_{k=1}^S \theta_k\right)t}\sum_{i=1}^S\frac{\theta_i}{\nu}}
{1+\sum_{k=1}^S \frac{\theta_k}{\nu}},\qquad
p_j(t)=\frac{\theta_j}{\nu}\frac{1-e^{-\left(\nu+\sum_{k=1}^S \theta_k\right)t}}
{1+\sum_{k=1}^S \frac{\theta_k}{\nu}},
\end{equation}
for $j=1,\dots, S$. It is easy to check that $0\le p_j(t) \le 1$ for $j=0,1,\dots,S$ and arbitrary time $t$. In addition, $\sum_{j=0}^S p_j(t)=1$ holds for any time. Therefore, the parameter combinations $p_j(t)$ can be interpreted as probabilities. As we immediately show below, they are precisely the probabilities associated to the $j$-th population in a \emph{multinomial} distribution. 

First, using the binomial theorem, we can expand $G(\bm{z},t)$ as
\begin{equation*}
G(\bm{z},t) = \biggl(p_0(t)+\sum_{j=1}^S p_j(t) z_j\biggr)^{n_A^0} 
= \sum_{n_A=0}^{n_A^0}\binom{n_A^0}{n_A}p_0^{n_A}\biggl(\sum_{j=1}^S p_jz_j\biggr)^{n_A^0-n_A}.
\end{equation*}
Then, by the multinomial theorem~\citep{knuth1997art}, it holds
\begin{equation*}
\biggl(\sum_{j=1}^S p_jz_j\biggr)^{n_A^0-n_A}=\!\!\!\!
\sum_{\substack{n_1,\dots,n_S\ge 0\\\sum_{i} n_i=n_A^0-n_A}}\!\!\!\!
\frac{(n_A^0-n_A)!}{n_1!\cdots n_S!}\prod_{k=1}^S (p_k z_k)^{n_k},
\end{equation*}
which substituted into the identity above yields
\begin{equation*}
P_{\bm{n}}(t)=\sum_{\substack{n_A=0\\n_A=n_A^0-\sum_{i} n_i}}^{n_A^0}
\frac{n_A^0!}{n_A!n_1!\cdots n_S!}p_0^{n_A}
p_1^{n_1}\cdots p_S^{n_S}
\end{equation*}
after identifying the sought probability $P_{\bm{n}}(t)$ as the coefficient of the series expansion associated with the power $z_1^{n_1}\cdots z_S^{n_S}$. Given the restriction $n_A=n_A^0-\sum_{i} n_i$ that affects the sum above, it is only one term that contributes to the sum, i.e., the one that verifies the restriction. Hence we obtain the definition~\eqref{eq:multidef} of the multinomial distribution. In addition, from the expression above it is apparent that if $\sum_{j=1}^S n_j > n_A^0$, then $P_{\bm{n}}(t)=0$. So we recover the multinomial distribution $\mathcal{M}(n_A^0,\{p_j(t)\})$ with $n_A^0$ trials and probabilities $p_j(t)$ given by Eq.~\eqref{eq:pj}.

This result agrees with those obtained at equilibrium, in the limit $t\to\infty$. As explained in~\ref{sec:appA00}, we also found a multinomial distribution in the stationary regime. However, the out-of-equilibrium distribution for $\nu_1=\nu_2=\dots=\nu_S$ is not multinomial in general: note that the generating function for non-zero initial populations, Eq.~\eqref{eq:Gtgen}, does not match the one one associated to a multinomial distribution, which is given by Eq.~\eqref{eq:Gt}, unless there are no handling predators initially. $P_{\bm{n}}(t)$ follows a multinomial distribution only if $\bm{\hat{n}}=\bm{0}$. Irrespective of the set of initial conditions, however, the generating function~\eqref{eq:Gtgen} has the correct limit as $t\to\infty$. Indeed, if we take this limit on both sides of Eq.~\eqref{eq:Gtgen} we get
\begin{equation*}
\lim_{t\to\infty} G(\bm{z},t)=\left(1-\frac{m({\bm{z}})}{\nu+Q}\right)^{n_A^0}
= \biggl(p_A+\sum_{j=1}^S p_j z_j\biggr)^{n_A^0},
\end{equation*}
where $p_A=\nu/(\nu+Q)$ and $p_j=\theta_j/(\nu+Q)$, which are precisely the probabilities given by Eqs.~\eqref{eq:pi} and~\eqref{eq:pA} in the particular case of equal relaxation rates. Therefore, in this limit we obtain the generating function~\eqref{eq:Gt} of the multinomial distribution $\mathcal{M}(n_A^0,\{p_A,p_1,\dots,p_S\})$ with $n_A^0$ trials. In sum, what we have shown here is that, in the case of equal handling times, and starting by an initial population formed solely by $n_A^0$ searching consumers, the joint probability distribution of all behavioral types is multinomial in the transient as well as at stationarity. Consistently, the probabilities $p_j(t)$ derived in the out-of-equilibrium regime, given by Eq.~\eqref{eq:pj}, satisfy $p_j(t)\to p_j$ ($j=1,\dots,S$) and $p_0(t)\to p_A$ in the limit $t\to\infty$, for $p_A$ and $p_j$ given by Eqs.~\eqref{eq:pi} and~\eqref{eq:pA}, respectively.

Out of equilibrium, feeding rates are perfectly identifiable, as can be seen from the generating function by computing an exhaustive summary. We assume that resource densities are known in advance and not subject to MLE, as well as the total number of consumers $n_A^0$. The generating function~\eqref{eq:Gt} is completely determined once the $S+1$ parameter combinations gathered in vector $\bm{\kappa}(\bm{\theta})=\left(\nu+\sum_{k=1}^S \theta_k,\frac{\theta_1}{\nu},\dots,\frac{\theta_{S}}{\nu}\right)$ are known, as can be seen immediately from Eq.~\eqref{eq:pj}. Therefore, the exhaustive summary determines uniquely the $S$ attack rates as $\alpha_i = \theta_i / x_{R_i}$ ($i=1,\dots,S$) and the reciprocal mean handling time, $\nu$. Thus, the model is structurally identifiable out of equilibrium.

\section{Effect of steady-state and out-of-equilibrium sampling}
\label{sec:ident}

The two solvable models discussed above (Sections~\ref{sec:gfun} and~\ref{sec:multi}) illustrate the issue of parameter redundancy when steady-state probability distributions are used: in that regime, any likelihood function based on steady-state distributions will allow only for precise estimation of ratios of rates, not the rates themselves. Therefore, we have to use time-dependent probability distributions to ensure structural identifiability. However, as opposed to structural issues in model design, limitations in the dataset often preclude the whole parameter set to be inferred from data ---this is known as \emph{practical identifiability}~\citep{wieland2021structural}. In this section we focus on the effect of the quality of the data: is the problem of identification failure fully solved if we retain the temporal dependence in the probability $P_{\bm{n}}(t)$, regardless of the times at which model samples are taken? 

To estimate parameters through maximum likelihood, we think of a procedure based on tracking population abundances in \emph{in silico} (simulated) model samples (we name this methodology as ``feeding experiment''). We want to estimate unique values for feeding rates based only on feeding experiments, in which one may have access to the population abundances of each behavioral type (in our case, the number of free consumers, $n_A$, and the number of consumers handling resource $i$, $n_i$, for $i=1,\dots,S$). We distinguish between two situations: (i) abundance data is available for different independent replicas of the system at different times at the steady state, and (ii) abundances for different behavioral types are available for out-of-equilibrium independent replicas of the system, taken at different finite times along the transient regime before reaching stationarity. We use the distributions computed in Section~\ref{sec:multi} to infer, if possible, unique values for attack rates and reciprocal mean handling times. 

In both cases, our {\em in silico} feeding experiments comprise $n_A^0$ individual consumers feeding on $S$ different resources at controlled resource densities ($x_{R_i}$ constant, $i=1,\dots,S$), under Holling type II stochastic dynamics ---cf. reaction schemes in Eqs.~(\ref{int_AR_Multi_HII})--(\ref{death_AR_Multi_HII}). The parameter values that we would like to estimate from abundance data are: attack rate $\alpha_i$ for each resource, as well as the reciprocal mean handling time over each resource, $\nu_i$ (relaxation rate). Assume a fixed system volume $N$. 
Suppose we have a detection device that can record the number of free consumers, $n_A(t_j)$, as well as the number of consumers handling on each resource $i$, $n_i(t_j)$ ($i=1,\dots,S$), at time $t_j$. In order to use the multinomial distribution~\eqref{eq:multidef} with probabilities~\eqref{eq:pj} as the probability of any configuration of the system, every replica of the experiment starts initially with $n_A(0)=n_A^0$ free consumers, and no consumer is handling any prey. From now on, for the sake of simplicity, we consider that all predator individuals have equal attack and relaxation rates across resources ($\alpha:=\alpha_1=\dots=\alpha_S$, $\nu:=\nu_1=\dots=\nu_S$). Therefore, only two parameters $\alpha$ and $\nu$ are to be estimated.

We replicate the experiment a number of times $T$, each replica simulating the stochastic dynamics (using~\cite{gillespie1977exact} method) with $n_A^0$ free predators initially, and record abundances at time $t_j$ for each realization $j=1,\dots,T$. As shown in~\ref{sec:appB}, it is important that at least two sampling times are different (i.e., $t_j\ne t_k$ for some $j\ne k$). Otherwise the model is not identifiable. In practice, from an empirical point of view, it is very unlikely that independent system samples are obtained at precisely the same exact time.

Thus, we have $T$ independent stochastic realizations of the system, each replica is an observation at a different time. The observational data is the following matrix
\begin{equation*}		      		
M(\{t_j\}) = \begin{pmatrix}
n_{1,1}(t_1) & n_{1,2}(t_2) & \cdots & n_{1,T}(t_T) \\
\vdots & \vdots & \ddots & \vdots \\
n_{S,1}(t_1) & n_{S,2}(t_2) & \cdots & n_{S,T}(t_T)
\end{pmatrix},
\label{eq:DATAMD2Dout}
\end{equation*}
where columns comprise vectors $\bm{n}_j=(n_{1,j},\dots,n_{S,j})$ reporting observed abundances of handling predators at sampling time $t_j$. Free predator abundance at time $t_j$ is computed with $n_{A}(t_j)=n_A^0-\sum_{i=1}^S n_{i,j}(t_j)$. If each column of $M(\{t_j\})$ is a sample of the Holling type II multi-resource dynamics, then the probability $P_{\bm{n}_j}(t_j\vert \bm{\theta})$, computed using Eqs.~\eqref{eq:multidef} and~\eqref{eq:pj}, is the probability of observing a population vector $(n_A(t_j),\bm{n}_j)$ given a model parameter vector, $\bm{\theta} := (\alpha,\nu)$ ---notice that resource densities, $x_{R_i}=n_{R_i}/N$, the system volume, $N$, and the total number of consumers, $n_A^0$, are kept constant by assumption and therefore are not regarded as parameters in $\bm{\theta}$. Matrix $M(\{t_j\})$ gathers $T$ independent observations, so the function that assigns a likelihood to parameter vector $\bm{\theta}$ given the data matrix $M(\{t_j\})$ is multiplicative, $\mathcal{L}(\bm{\theta}\vert M(\{t_j\})) = \prod_{j=1}^{T} P_{\bm{n}_j}(t_j\vert \bm{\theta})$. The log-likelihood function, 
\begin{equation}
\log \mathcal{L}(\bm{\theta}\vert M(\{t_j\})) = \sum_{j=1}^{T} \log P_{\bm{n}_j}(t_j\vert \bm{\theta}),
\label{eq:logLKMD2D}
\end{equation}
is more convenient for numerical purposes.

\begin{figure}[t!]
\centering
\includegraphics[width=0.6\columnwidth]{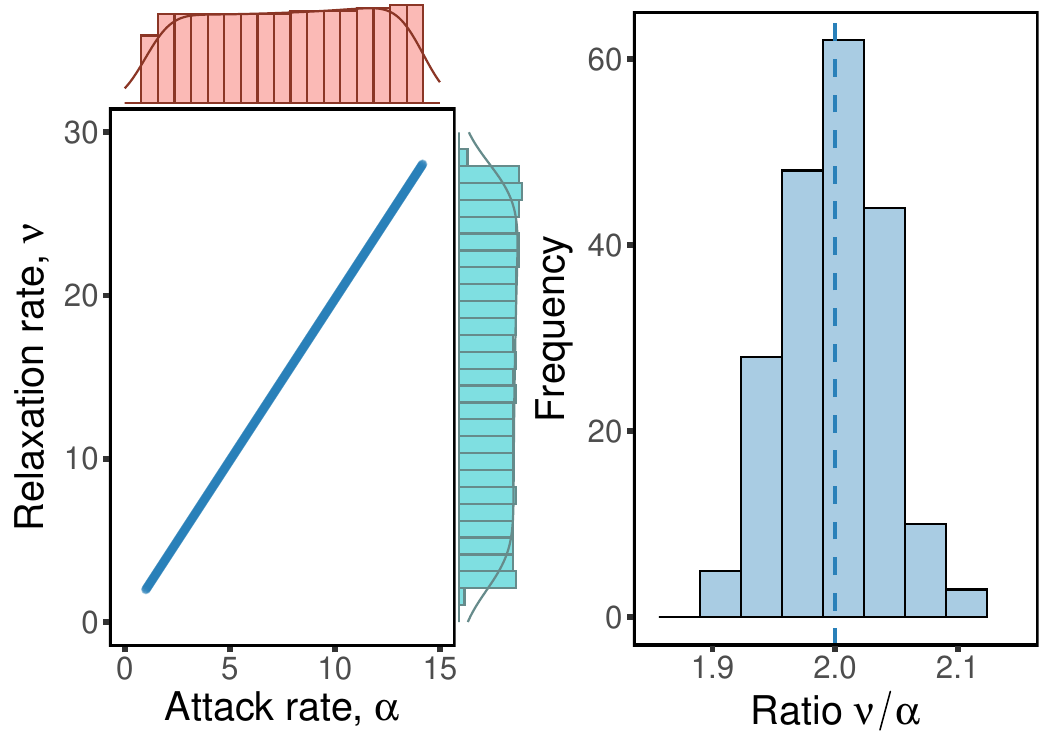}
\caption{\label{fig:eq}\textbf{Estimation using equilibrium samples reveals parameter redundancy}. We generated a single realization of the data matrix $M(\{t_j\})$, including $T=100$ independent model realizations starting from $n_A^0=100$ searching predators at $t=0$, using Gillespie's stochastic simulation algorithm up to $T=100$ different ending times uniformly drawn in the interval $[50\tau,100\tau]$ to ensure stationarity, with $\tau=(\nu+\alpha\sum_{i=1}^S x_{R_i})^{-1}$ the characteristic time of the dynamics. True rates are $\alpha=5$, $\nu=10$, and  individual consumers feed on $S=5$ resources with densities $\bm{x_R}=(0.9,0.7,0.5,0.3,0.1)$. Left: for that synthetic dataset, we run a L-BFGS-B optimization algorithm for the log-likelihood function~\eqref{eq:logLKMD2D} starting from different initial conditions, and we recorded the values of $(\alpha^{\star},\nu^{\star})$ at which the maximum was reached. Optimal parameter values follow a perfect straight line. The distributions of $\alpha^{\star}$ and $\nu^{\star}$ are compatible with uniform distributions, and the optimal values of $\log\mathcal{L}$ keep constant up to round-off errors. Overall, this indicates a line of infinite optima. Right: we replicated the previous experiment 200 times and, for each replica, we estimated the ratio $\nu/\alpha$ by fitting the slope of the straight line. We obtained the distribution in the right panel, which is compatible with a Gaussian with mean 2 and variance extracted from data. The vertical, dashed line indicates the true ratio.} 
\end{figure}

\begin{figure*}[t!]
\centering
\includegraphics[width=\textwidth]{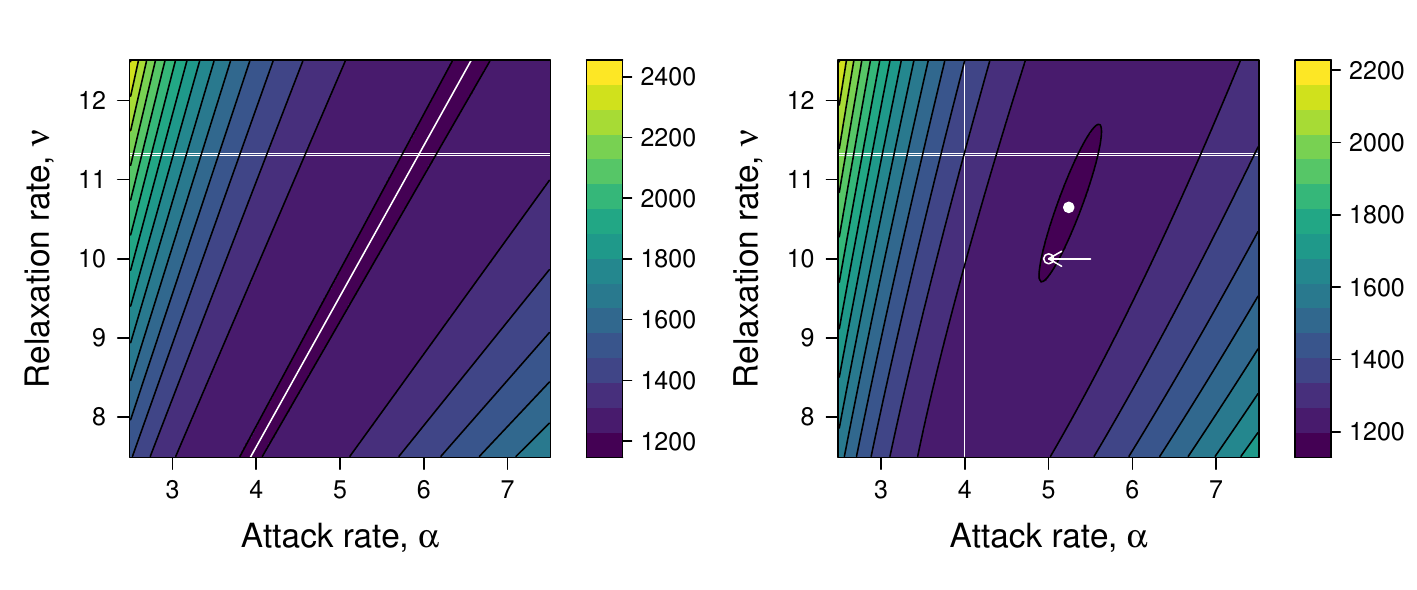}
\caption{\label{fig:heat} \textbf{Heatmaps of negative log-likelihood for stationary and out-of-equilibrium samples unveils redundancy} (for steady-state sampling). Representation of $-\log\mathcal{L}$, cf. Eq.~\eqref{eq:logLKMD2D}, in the $(\alpha,\nu)$ space, obtained for two sets of $T=100$ model samples generated for $(\alpha,\nu)=(5,10)$: (i) taken at stationarity (sampling times $\{t_j\}$ drawn uniformly in $[10\tau,20\tau]$, left panel), and (ii) taken out of equilibrium (sampling times $\{t_j\}$ drawn uniformly in $[0,10\tau]$, right panel), $\tau$ being the characteristic time scale. Left: a line of infinite minima is obtained. Right: The arrow marks the true pair $(\alpha,\nu)=(5,10)$ used for data generation, and the white point marks the minimum of $-\log\mathcal{L}$. In both panels, remaining parameter values are the same as in Fig.~\ref{fig:eq}.} 
\end{figure*}

Now, consider the effect of sampling times and assume that matrix $M(\{t_j\})$ is obtained when $t_j$ are ``large enough'' compared to the characteristic time of the dynamics, which is given by $\tau=(\nu+\alpha\sum_{i=1}^S x_{R_i})^{-1}$ ---see Eq.~\eqref{eq:pj}. In Fig.~\ref{fig:eq}, \emph{in silico} data (i.e., abundances of each behavioral type) were generated by stochastic simulation of a system with $S$ resources (with fixed densities $x_{R_i}$) consumed by $n_A^0$ predator individuals. A number $T$ of independent stochastic simulations started with  initial conditions $n_A=n_A^0$ and $n_i=0$ ($i=1,\dots,S$) at $t=0$, each of which were stopped up to a time $t_j$ drawn uniformly in $[50\tau,100\tau]$, which is enough to reach stationarity. We maximized the log-likelihood function~\eqref{eq:logLKMD2D} using the L-BFGS-B optimization scheme~\citep{zhu1997algorithm}, which was used for numerical optimization throughout the contribution. Each realization of the data matrix $M(\{t_j\})$ leads to a line of maximum likelihood in the $(\alpha,\nu)$ space (see also Fig.~\ref{fig:heat}, left panel), whose slope can vary over different replicas but can be determined precisely, providing to a consistent, unbiased estimation of the ratio $\nu/\alpha$ (Fig.~\ref{fig:eq}, right panel). Using data sampled at stationarity, even if ML is based on a time-dependent log-likelihood, only parameter ratios are identifiable.

From Eqs.~\eqref{eq:n_i_star_HII_multi}--\eqref{eq:n_A_star_HII_multi} in \ref{sec:appA00}, it is immediate to check that the dimensionless ratio $\nu/\alpha$ satisfies
\begin{equation}\label{eq:slope}
\frac{\nu}{\alpha}=\frac{\langle n_A\rangle}{\langle n_{AR}\rangle}\sum_{i=1}^S x_{R_i},
\end{equation}
with $\langle n_{AR}\rangle := \left\langle \sum_{i=1}^S n_i \right\rangle =
\sum_{i=1}^S \langle n_i \rangle$ the average total number of handling consumers. In~\ref{sec:appB} we show that, when sampling times coincide ($t_1=t_2=\dots=t_T=t$) and $t\to\infty$, the likelihood function does exhibit a line of infinite critical points with slope given by~\eqref{eq:slope}, in which quantities $\langle n_A\rangle$ and $\langle n_{AR}\rangle$ are substituted by averages across samples, $\langle n_A\rangle_T := \frac{1}{T}\sum_{k=1}^T n_{A,k}$ and $\langle n_{AR}\rangle_T := \frac{1}{T}\sum_{k=1}^T \sum_{i=1}^S n_{i,k}$, respectively. Therefore, our numerical experiment (Fig.~\ref{fig:eq} and Fig.~\eqref{fig:heat}, left panels) is consistent with a submanifold of degenerated log-likelihood optimum values in the parameter space. 

\begin{figure*}[t!]
\centering
\includegraphics[width=\textwidth]{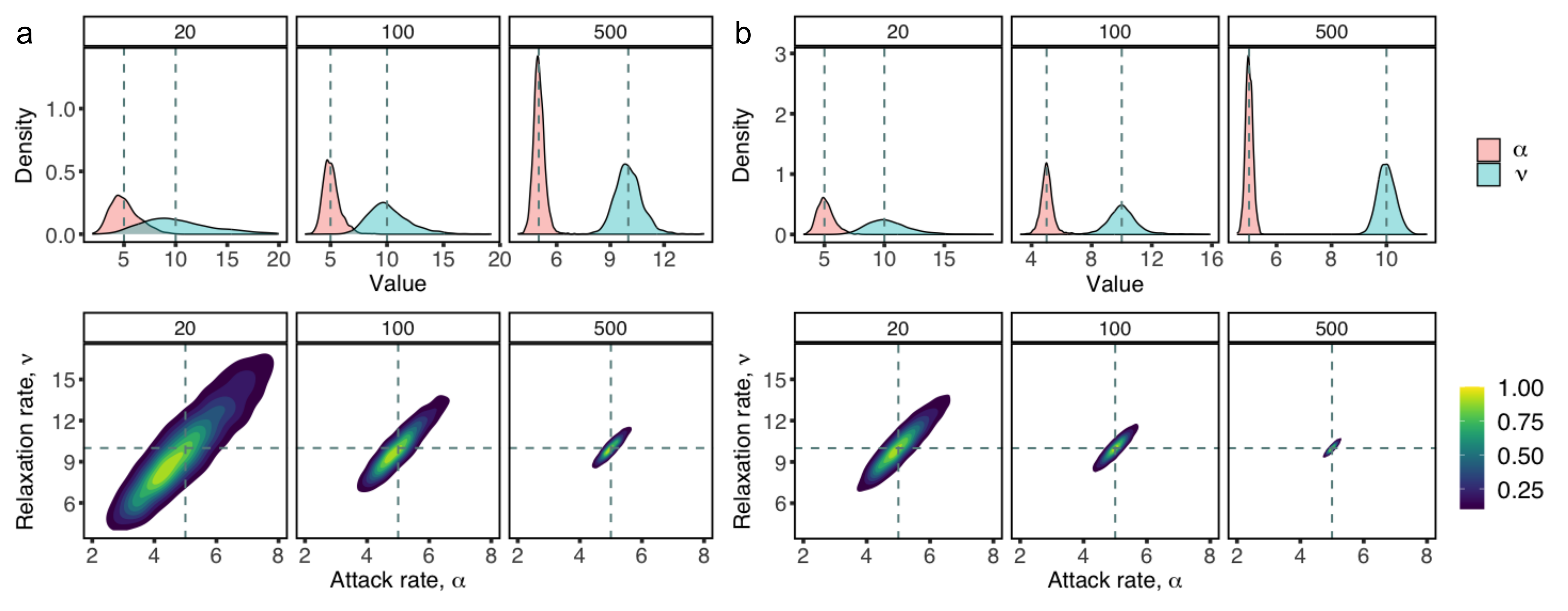}
\caption{\label{fig:dist} \textbf{Distributions of ML parameter estimates are unbiased for different number of initial consumers and increased number of temporal samples}. (a) The estimation procedure was repeated three times, for $n_A^0\in \{20,100,500\}$, using $T=20$ samples in simulated out-of-equilibrium data. Upper panels show the distributions of $\alpha^{\star}$ and $\nu^{\star}$, which tend to concentrate around true values (dashed lines) as $n_A^0$ increases. Parameter estimates show bias for low $n_A^0$ values, and the bias reduces as $n_A^0$ increase. This phenomenon is more clearly observed in the density plots represented in lower panels for $(\alpha^{\star},\nu^{\star})$, the lighter the color indicating the higher density of estimated pairs. (b) Same as (a) for $T=100$ temporal samples. Compared to panel (a), we observe that variability is decreased as $T$ increases. In both panels, remaining parameters are the same as in Fig.~\ref{fig:eq}. $3000$ estimation experiments were conducted to obtain distributions.} 
\end{figure*}


ML estimation techniques produce proper estimates for individual rates $\alpha$ and $\nu$ when data sample is taken out of equilibrium (and, importantly, at least two sampling times are different; see~\ref{sec:appB}). During the transient regime, samples were taken at times $\{t_j\}$ drawn from a uniform distribution over the interval $[0,10\tau]$, $\tau=(\nu+\alpha\sum_{i=1}^S x_{R_i})^{-1}$ being the characteristic dynamical time scale. Figure~\ref{fig:heat} compares the log-likelihood landscapes obtained using steady-state (left panel) and out-of-equilibrium model samples (taken at times uniformly drawn over the interval $[10\tau,20\tau]$). A parameter screening in the $(\alpha,\nu)$ plane shows a line of infinite maxima $\nu^{\star}\propto \alpha^{\star}$ at equilibrium, confirming the results reported in Fig.~\ref{fig:eq}. This evidences parameter redundancy when out-of-equilibrium samples are missing. On the contrary, a single log-likelihood maximum value arises and parameters can be truly inferred from the data when out-of-equilibrium samples are included (we have checked that, if samples are taken at times within the interval $[0,20\tau]$, we also find a single maximum in log-likelihood, which implies that redundancy is avoided even when the data contain steady-state and out-of-equilibrium observations together). The accuracy of the estimates $\alpha^{\star}$ and $\nu^{\star}$ depends on data quality. For example, if the number of samples $T$ taken is large, then it is expected that estimates will be close to actual values. Similarly, if the overall number of individuals $n_A^0$ monitored in the experiment is large, then better estimates for model parameters are expected. 


This phenomenon is illustrated in Figure~\ref{fig:dist}, which shows the distributions of ML estimates $\alpha^{\star}$ and $\nu^{\star}$ for $T=20$ and $T=100$ samples, and for three values of $n_A^0$. We observe that distributions become closer to true values as $T$ and $n_A^0$ increase. The joint distributions of parameter estimates (bottom panels) show also that estimates are consistent: although for small samples and low values of $n_A^0$ the joint distributions are asymmetric and not centered in true values and show certain bias, as $T$ and $n_A^0$ increase the distributions tend to center around actual parameter values with small variability.

These distributions allow to calculate $95\%$ confidence intervals for parameter estimates. We  calculated the $2.5\%$ and $97.5\%$ percentiles of the empirical distributions obtained in Fig.~\ref{fig:dist} (upper panels), these values corresponding to the limits of the sought confidence interval. Intervals are listed in Table~\ref{tab:CI} for the set of $T$ and $n_A^0$ values reported in Fig.~\ref{fig:dist}. As described, intervals are not centered in actual values for small samples, and tend to be centered (with decreasing spread) as $T$ and $n_A^0$ increase.

\begin{table}
\centering
\begin{tabular}{|c|c|c|c|}
\multicolumn{4}{c}{$95$\% Confidence Intervals} \\
\hline
$n_A^0$ & True value & $T=20$ times & $T=100$ times \\
\hline
20  & $\alpha=5$ & [2.86,8.25] & [3.95,6.81] \\
100 & $\alpha=5$ & [3.86,6.89] & [4.22,6.04] \\
500 & $\alpha=5$ & [4.47,5.69] & [4.75,5.28] \\
\hline
20  & $\nu=10$ & [4.86,17.76] & [7.42,14.32] \\
100 & $\nu=10$ & [7.25,14.55] & [8.08,12.54] \\
500 & $\nu=10$ & [8.67,11.66] & [9.40,10.67] \\
\hline
\end{tabular}
\caption{\textbf{Confidence intervals calculated using parameter distributions}. Here we present true values and $95\%$ confidence intervals for the {\em in silico} experiments shown in Fig.~\ref{fig:dist}. The empirical distributions of parameter estimates, shown in that figure, are used to compute the corresponding percentiles that form the confidence interval. The sample of estimated parameters used to compute intervals has a size of $3000$ estimates for each $(T,n_A^0)$ combination.}
\label{tab:CI}
\end{table}

\begin{figure}[t!]
\centering
\includegraphics[width=0.6\columnwidth]{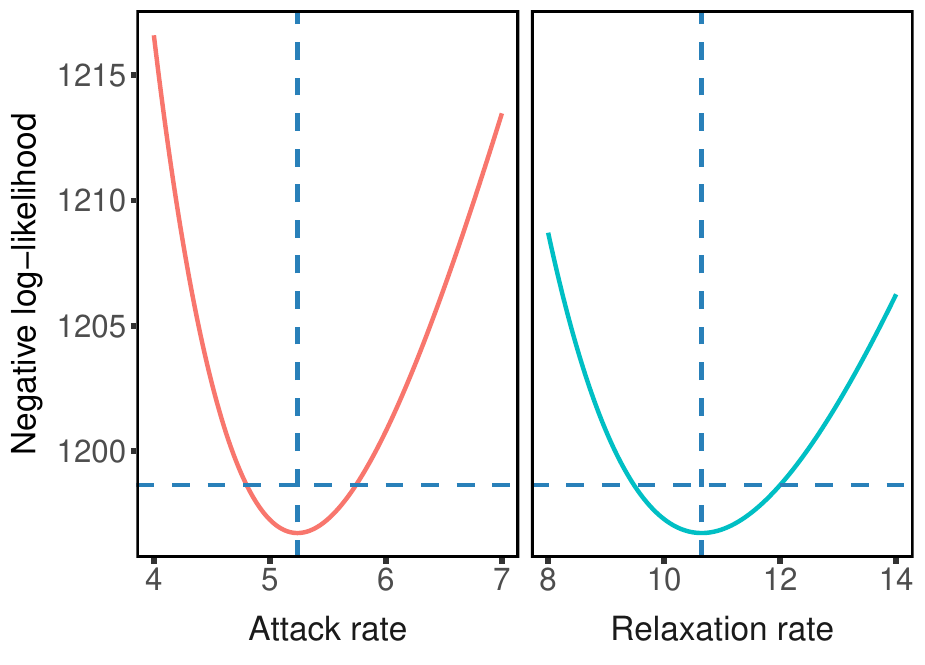}
\caption{\label{fig:profiles} \textbf{Confidence interval estimation using negative log-likelihood profiles} for the same synthetic matrix $M(\{t_j\})$ used for the heatmap in Fig.~\ref{fig:heat} (right panel). Negative log-likelihood profiles $\ell_1(\alpha)$ and $\ell_2(\nu)$ (left and right panels, respectively) yield confidence intervals as the abscissae of the intercept of the profiles $\ell_1(\alpha)$ and $\ell_2(\nu)$ with the horizontal dotted lines whose ordinates are equal to $\ell_1(\alpha^{\star})+\chi^2_{1,0.95}/2$ and $\ell_2(\nu^{\star})+\chi^2_{1,0.95}/2$, respectively. $(\alpha^{\star},\nu^{\star})$ refers to the point (marked with vertical, dotted lines) at which the maximum of negative log-likelihood is reached.}
\end{figure}

Usually, one can only have access to one single realization of empirical observations. In this case, we still can obtain ML parameter estimates as well as confidence intervals following the log-likelihood ratio test~\citep{cox1989analysis}. We illustrate this procedure using the same data matrix as in Fig.~\ref{fig:heat} (right panel), for which the ML parameter estimates are $(\alpha^{\star},\nu^{\star})=(5.24,10.65)$. We obtained log-likelihood profiles shown in Fig.~\ref{fig:profiles} by minimizing negative log-likelihood over $\nu$ for fixed $\alpha$, $\ell_1(\alpha):=-\min_{\nu}\log\mathcal{L}(\alpha,\nu\vert M(\{t_i\}))$ (left panel), and minimizing over $\alpha$ for fixed $\nu$, $\ell_2(\nu):=-\min_{\alpha}\log\mathcal{L}(\alpha,\nu\vert M(\{t_i\}))$ (right panel). Then, the $95\%$ confidence interval for $\alpha$ can be defined as the set of values satisfying $\vert \ell_1(\alpha)-\ell_1(\alpha^{\star})\vert \le \chi^2_{1,0.95}/2$, $\chi^2_{1,0.95}$ being the $95\%$ percentile of the chi-squared distribution with one degree of freedom. Similarly, the $95\%$ confidence interval for $\nu$ is defined as the set of values satisfying $\vert \ell_2(\nu)-\ell_2(\nu^{\star})\vert \le \chi^2_{1,0.95}/2$. Note that $\ell_1(\alpha^{\star})=\ell_2(\nu^{\star})$ equals the minima obtained for the negative log-likelihood at the point $(\alpha^{\star},\nu^{\star})$ as well. This yields the intervals $\alpha\in [4.80,5.73]$ and $\nu\in [9.51,11.98]$ at a $95\%$ confidence level, which are consistent with their counterpart in Table~\ref{tab:CI} for $T=100$ and $n_A^0=100$.

\begin{figure*}[t!]
\centering
\includegraphics[width=\textwidth]{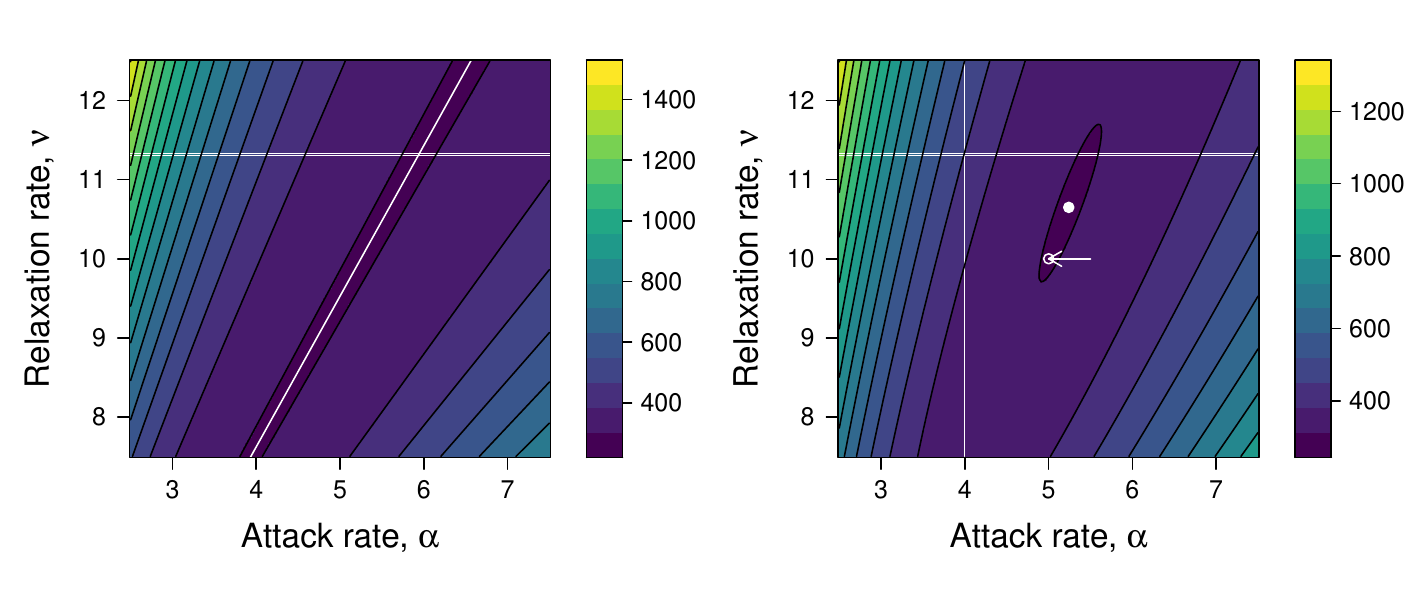}
\caption{\label{fig:heat_binom} \textbf{Heatmaps of negative log-likelihood using only free predator abundances for stationary and out-of-equilibrium samples unveils redundancy} (for steady-state sampling). Representation of $-\log\mathcal{L}$, cf. Eq.~\eqref{eq:LKHII} (i.e., using only free predators abundance information), in the $(\alpha,\nu)$ space, obtained for the same two sets of model samples generated for Fig.~\ref{fig:heat}: true parameter values are $(\alpha,\nu)=(5,10)$ (arrow in the right panel), and stationary (out-of-equilibrium) samples yield the log-likelihood represented in the left (right) panel. Observe that the heatmaps only differ from those in Fig.~\ref{fig:heat} in the colorbar values, because here we are using only partial information to evaluate $-\log\mathcal{L}$ for the same model observations as in that figure. Again, we find parameter redundancy for stationary sampling, whereas a single minimum is obtained for out-of-equilibrium sampling (the white point marks the minimum of $-\log\mathcal{L}$). Remaining parameter values were chosen as in Fig.~\ref{fig:eq}.} 
\end{figure*}

In case the experimental detection device could only count the number of free consumers at each sampling time, then we should resort on the one-dimensional marginal probability distribution of free consumers to build the correct likelihood function for our estimation procedure. Because any marginal of a multinomial is again a multinomial distribution (in particular, a binomial distribution if all variables but one are marginalized), the marginal distribution of the number of free predators is given by the binomial~\eqref{eq:binom} with
\begin{equation*}
p_A(t)=\frac{\nu+\alpha\, e^{-\left(\nu+\alpha \sum_{k=1}^S x_{R_k}\right)t}\sum_{i=1}^S x_{R_i}}
{\nu+\alpha\sum_{k=1}^S x_{R_k}},
\end{equation*}
as obtained in Eq.~\eqref{eq:pj}. Then we can maximize the log-likelihood function
\begin{equation}
\mathcal{L}(\bm{\theta}\vert M(\{t_j\})) = \prod_{j=1}^{T} P_{n_{A,j}}(t_j\vert \bm{\theta})
\label{eq:LKHII}
\end{equation}
where, again, independence has been assumed in stochastic realizations, and $\bm{\theta}=(\alpha,\nu)$ represents the vector of varying model parameters. Here 
\begin{equation*}
M(\{t_j\})=(n_{A,1}(t_1), n_{A,2}(t_2), \dots, n_{A,T}(t_T))
\label{eq:DATAMD1Dout}
\end{equation*}
stands for an empirical data vector with $T$ observational abundances $n_{A,j}$ of the number of free predators. We use the time-dependent log-likelihood to estimate $\alpha$ and $\nu$ depending on the sampling scheme used to generate the vector $M(\{t_j\})$. The ratio $\nu/\alpha$ can be inferred if the experiment only reports steady-state abundance data, but not rates themselves. Model identifiability is illustrated with the plots of log-likelihood out-of and at equilibrium in Fig.~\ref{fig:heat_binom}, where only searching predators abundance data is used for ML parameter estimation. As in the first experiment, in which the abundances of each behavioral type were available, unique parameter estimates can be inferred from out-of-equilibrium model observations. 

\begin{figure*}[t!]
\centering
\includegraphics[width=\textwidth]{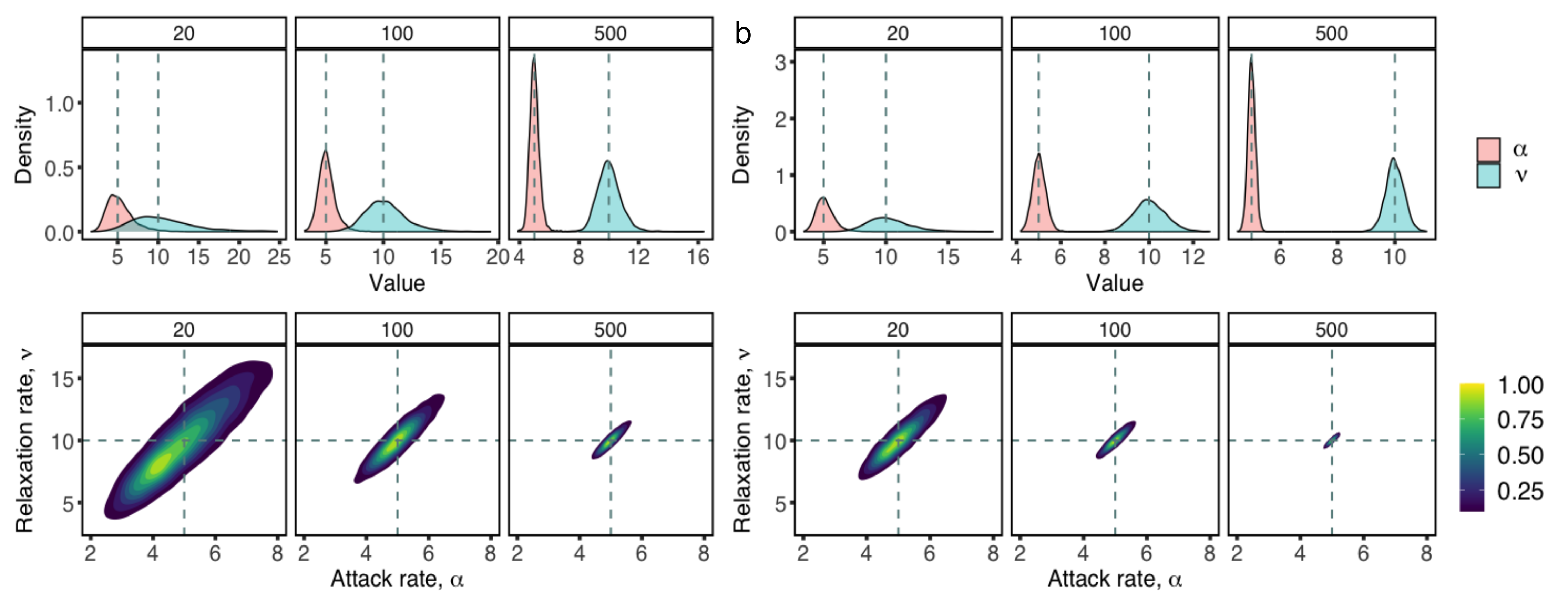}
\caption{\label{fig:dist_binom} \textbf{Distributions of ML parameter estimates using only free predator abundances are unbiased for increasing number of initial consumers and number of temporal samples}. Same as Fig.~\ref{fig:dist} for a ML estimation procedure based on searching predator numbers. In this case, we have only used the number of free predators recorded at each sampling time, see Eq.~\eqref{eq:DATAMD1Dout}. Distributions are quite similar to those obtained by using the vector of population handling predator abundances, $\bm{n}=(n_1,\dots,n_S)$, which are shown in Fig.~\ref{fig:dist}. In both panels, parameters are the same as in Fig.~\ref{fig:eq}.} 
\end{figure*}

Parameter estimation is illustrated with a second {\em in silico} experiment based only on free predators abundance vectors ---cf. Eq.~\eqref{eq:DATAMD1Dout}. Figure~\ref{fig:dist_binom} shows the distributions obtained in this case, gathering $3000$ ML parameter estimates for varying $n_A^0$, the overall number of consumers, and $T$, the number of observations (columns) used to construct vector $M(\{t_j\})$. Distributions for each parameter (upper panels) are very similar to those reported in Fig.~\ref{fig:dist}, which took information of the full vector of abundances. Only for small-size samples (i.e., for $n_A^0$ and $T$ small), distributions and density plots (lower panels) appear to be more spread in this case. This effect can be seen more clearly in confidence intervals, which are calculated using the empirical distributions in Fig.~\ref{fig:dist_binom}: reported intervals (Table~\ref{tab:CI_binom}), based only on free predator abundances, are wider than those listed in Table~\ref{tab:CI}, for $n_A^0=20$ and $T=20$. Once the ML estimation procedure contains more information, as $n_A^0$ and $T$ increase, the intervals estimated from both synthetic experiments are consistently comparable. 

\begin{table}
\centering
\begin{tabular}{|c|c|c|c|}
\multicolumn{4}{c}{$95$\% Confidence Intervals} \\
\hline
$n_A^0$ & True value & $T=20$ times & $T=100$ times \\
\hline
20  & $\alpha=5$ & [2.93,9.34] & [3.95,6.80] \\
100 & $\alpha=5$ & [3.82,6.94] & [4.47,5.64] \\
500 & $\alpha=5$ & [4.44,5.71] & [4.75,5.25] \\
\hline
20  & $\nu=10$ & [4.81,20.27] & [7.36,14.40] \\
100 & $\nu=10$ & [7.12,14.62] & [8.69,11.54] \\
500 & $\nu=10$ & [8.64,11.75] & [9.37,10.63] \\
\hline
\end{tabular}
\caption{\textbf{Confidence intervals using only free predators abundances} calculated using parameter distributions (Fig.~\ref{fig:dist_binom}). The table lists actual values and $95\%$ confidence intervals for parameters of the {\em in silico} experiments shown in Fig.~\ref{fig:dist_binom}. Again, $3000$ samples of estimated parameters were used to compute intervals for each $(T,n_A^0)$ combination.}
\label{tab:CI_binom}
\end{table}

\section{Discussion}
\label{sec:discussion}

In this contribution we have studied model identifiability in two stochastic population models, and we have shown that, when independent model observations include samples taken at different sampling times in the transient regime, the model is truly identifiable in the sense that model parameters can be uniquely estimated from observations. On the contrary, if population numbers are only observed at stationarity, then the model is set-identifiable because only a combination of parameters can be inferred from data. We derived model redundancy conditions using the generation function formalism in both a single-popultion birth-death-immigration model, and a multi-resource consumer-resource stochastic population model. For the latter, synthetic data were used to calibrate the model in two \emph{in silico} experiments: one in which only free predator population numbers are considered, and another in which the whole matrix of searching and handling predator numbers is available. If this model was to be calibrated using actual, experimental data, and independent observations were taken at stationarity (or at equal sampling times), the problem of parameter redundancy would persist: the log-likelihood function would produce an infinite set of maxima, because the critical points conditions would reduce to one single equation ---see the derivation of Eq.~\eqref{eq:mani} in~\ref{sec:appB}. 

Importantly, if temporal samples contain both transient and stationary observations, model parameters remain perfectly identifiable. Therefore, an important implication of our work is that time series must include out-of-equilibrium observations to infer parameters without redundancy. Several statistical methods allow one to separate transient and steady-state regimes in a time series: for example, the Chow test~\citep{chow1960tests} allows to separate a positive slope in abundance time series from a slope compatible with zero, proper of a steady-state regime. For empirical data, the experiment has to be designed to collect out-of-equilibrium samples, by monitoring the experiment from an initial condition far from equilibrium.

For stochastic processes other than the multi-resource consumer-resource model analyzed here, we can state that, as long as the distributions of observational variables depend on parameter combinations whose intersection is a manifold that does not reduce to a unique point, then identification failure will emerge. This is particularly important at equilibrium, because many models are calibrated using steady-state distributions (which are the ones commonly available, unlike the out-of-equilibrium distributions, which are difficult to derive analytically). If those steady-state distributions depend on parameter ratios, as in the examples discussed here, then the issue of parameter inference will arise. Because many quantities at equilibrium depend on dimensionless parameter combinations, we anticipate that the problem of parameter redundancy can be very frequent in stochastic models commonly used. This has important implications on ML parameter estimation using empirical data, because if the system under study is sampled once the steady-state has been reached, the problem of parameter redundancy is likely to arise.

Additional information should be available to conduct precise parameter inference at stationarity. The heatmaps in Fig.~\ref{fig:heat} represent the log-likelihood function values, Eq.~\eqref{eq:logLKMD2D}, as the two parameters vary between a minimum and a maximum value along the axes, and show that when samples are taken close enough to stationarity, then a line of optimal values arise. This agrees with the expression found in~\ref{sec:appA00} for the joint stationary distribution (as well as its marginals), which depends on the compound parameter controlled by the dimensionless ratio $\nu/\alpha$ (the relaxation rate in units of the attack rate). This nondimensional quotient is the one that can be statistically inferred from data at equilibrium. If we were able to sort out, experimentally, individuals in the different behavioral modes at stationarity (attacking {\em vs} handling), we could identify, as shown in Figure~\ref{fig:heat}, the true attack rate $\alpha$ on resources, respectively, provided that we had external information to previously determine the relaxation rate $\nu$, and confidence intervals could be assigned (see Table \ref{tab:CI}). Regarding to the rest of parameters, that is, the total number of consumers, $n_A^0$, resource densities $x_{R_i}$, and the volume, $N$, of the system, we believe that they will be fairly easy to initially fix and control in the setting of any experiment.

We stress that any additional information used to solve parameter redundancy at equilibrium should come from monitoring the system out of equilibrium. In particular, precise estimates of $\tau$ would break redundancy at stationarity. Observe that, if the characteristic time $\tau$ and the equilibrium distribution are known, one can obtain non-redundant parameter estimates in the case of equal handling times. However, estimating the characteristic time requires knowledge of the transient dynamics before reaching the steady state. One could perturb an experimental system by varying populations along a proper direction away from the equilibrium point (along an eigenvector of the Jacobian matrix evaluated at equilibrium). Then the eigenvalue would help measure the relaxation time. Methods alternative to our ML approach, such as the one proposed in~\ref{sec:appC} (based on the initial decay of the abundance of free predators), imply breaking stationarity or having precise knowledge of the transient state. These methods are, therefore, equivalent to sampling the system during the transient dynamics before stationarity.


As for the generality of our results, the multinomial distribution derived in section~\ref{sec:multi}, either at stationarity or out of equilibrium, can be extended to a situation in which predator's feeding efficiency varies non-linearly in response to both the same or different resources. Let $\alpha_{ij}$ ($i,j=1,\dots,S$) measure the effect of resource densities $x_{R_i}$ and $x_{R_j}$ on the formation of consumers handling on resource $i$. The effect of the density of the same resource or alternative resources on the formation of handling consumers of the focal resource $i$ can be taken into account, for example, if we assume the following $3\times S$ chemical reaction scheme:
\begin{eqnarray}
X_A + 2\,X_{R_i}           \ch{&->[ $\alpha_{ii}$ ]} & X_{AR_i} + X_{R_i},
\label{int_AR_ii_HIII}\\
X_A +   X_{R_i} + X_{R_j}  \ch{&->[ $\alpha_{ij}$ ]} & X_{AR_i} + X_{R_j}, \label{int_AR_ij_HIII}\\
X_{AR_i} \ch{&->[ $\nu_{i}$ ]} & X_A, \label{death_AR_i_HIII}
\end{eqnarray}
where the second reaction holds for $i\ne j$. In this way, depending on the balance between $\alpha_{ii}$ and $\alpha_{ij}$, consumer's feeding rate on resource $R_i$ can be enhanced or damped by the presence of an alternative resource $R_j$ more than it is by the presence of the same type of resource $R_i$. Therefore, consumer efficiency on a focal resource is affected not only by the density of the focal resource but also by the presence of alternative resources. 

Proceeding in the same way as in~\cite{palamara2022implicit}, the functional response that arises from those fundamental processes corresponds to a multispecies version of the Holling type III functional response. It is easy to see that, according to the reaction scheme~\eqref{int_AR_ii_HIII}--\eqref{death_AR_i_HIII} and because of mass action, the per capita consumption rate of a free/searching individual $X_A$ feeding on a focal resource $i$ becomes
\begin{equation}
\theta_i := \alpha_{ii} (x_{R_i}^0)^2 + \sum_{j\ne i} \alpha_{ij}\,x_{R_i}^0\,x_{R_j}^0.
\label{eq:theta_Multi_i}
\end{equation} 
If only a single resource were present, the first term would correspond to a Holling type III functional response, with Hill exponent (i.e., the exponent to which resource densities are raised) $r=2$~\citep{real1977}. The definition of per-capita consumption rates $\theta_i$ would be the only change in the derivation of the distributions, at stationarity or out of equilibrium, for this scheme of elemental processes. All the results derived for the multi-resource Holling type II case treated in section~\ref{sec:multi} hold in this more general case, because transition probability rates are exactly the same except for the replacement of parameter $\theta_i$ by the one in Eq.~\eqref{eq:theta_Multi_i}, which is still a constant parameter that depends on resource densities under the hypothesis of chemostatic conditions. Indeed, note that the two channels specified by Eqs.~\eqref{int_AR_ii_HIII} and~\eqref{int_AR_ij_HIII} can be subsumed into the transition probability rate for increasing the number of handling predators by one unit ---first equation of ~\eqref{eq:rates}, with $\theta_i$ given by~\eqref{eq:theta_Multi_i}. Reaction~\eqref{death_AR_i_HIII} leads to the same transition rate for decreasing the number of handling predators by one unit ---second equation of~\eqref{eq:rates}. As a consequence, the previous derivation of the multi-resource Holling Type II stationary and out-of-equilibrium distributions applies here replacing parameter $\theta_i$ by the new definition. In both type II and III cases, the formulae that define probability parameters of the multinomial distributions remain valid with the replacement of $\theta_i$ by its expression in~\eqref{eq:theta_Multi_i}. Consequently, the experiments designed to carry out ML parameter estimation can be generalized to the Holling type III dynamics as well, together with the issues of parameter identifiability. Our results can be extended to more general consumer-resource contexts, unveiling the issue of parameter redundancy to be very common in stochastic population models.


We used the assumption of equal handling times to obtain closed-form solutions for the out-of-equilibrium joint distribution of population abundances. When handling times for different resources are not equal, our methodology applies if the master equation is integrated numerically to obtain probability distributions as functions of time. Given these probabilities, one can compute numerically log-likelihoods and estimate the $2S$ parameters $\{\alpha_i,\nu_i\}$ ($i=1,\dots, S$) without redundancy. It is important to note that, in this case, the degeneracy of the approach when samples are taken at equilibrium increases because, in this case, only $S$ ratios ($\alpha_i/\nu_i$) are identifiable ---see Eqs.~\eqref{eq:pi}--\eqref{eq:pA}. If only abundance data at stationarity is available, to solve redundancy one would need prior knowledge of $S$ additional quantities relating the unknown rates.
On a side note, it is important to observe that the assumption of equal handling times does not impose any restriction in the case of one single resource: for the standard Holling type II consumer-resource feeding dynamics, the calculations that follow are fully general for the case $S=1$, and the multinomial out-of-equilibrium solution is exact in that case ---provided that the dynamics starts with $n_A^0$ searching consumers initially.

Traditional methods for estimating attack and handling rates in consumer-resource interactions rely on depletion experiments, where a fixed number of consumers are placed with a known initial resource density and the abundance of resources decreases over time due to consumption~\citep{rogers1972random}. Parameters like attack rate and handling time are inferred by fitting the depletion curve to a model of resource abundance. These experiments run out-of-equilibrium, without resource replenishment. In contrast, our approach assumes chemostatic conditions, where consumed resources are continuously replenished, and the total number of consumers remains constant. By tracking consumer behavior ---distinguishing between searching and handling states--- through animal trajectories, one could apply our maximum likelihood approach to estimate process rates. Monitoring the early stages of the experiment is crucial to avoid parameter redundancy. Experimental setups capable of classifying individual behaviors, such as searching for or digesting resources, can support this approach and allow parameter estimation under controlled conditions~\citep{Bartumeus02,Bartumeus05,Bartumeus03}. Because maximum likelihood estimators are asymptotically those with minimal variance, we expect that our approach is better than simply fitting a curve to data, as in depletion experiments.

Another shortcoming of the inference approach followed in this contribution comes from the independence of measurements at different times, needed to construct the likelihood function to uniquely estimate feeding rates. Such assumptions would require the use of different replicates for every measurement, which would importantly increase the cost of hypothetical field or laboratory experiments due to the repetition of realizations. In principle, one could not realistically use the same replica to gather different measurement at different times, because the independence hypothesis would be broken, together with sampling the system close to equilibrium as time elapses. Other inference approaches, not based on multiplicative likelihood functions, should be devised to extract meaningful parameter estimates from controlled experiments. 

Classical feeding experiments have been performed for a variety of systems and taxa, where feeding rates have been measured mainly using the averaged expressions at fixed resource levels~\citep{novak2020systematic,stouffer2021hidden,delong2021predator}. Here, we propose to use the distribution of behavioral types, thus allowing for refined estimates of parameters associated to elemental process. For the multi-resource Holling type II stochastic dynamics (as well as the extensions discussed here), we managed to give a complete analysis of the temporal dynamics, as well as its steady state. We described the potential inference methods to use to estimate, when identifiable, population parameters or characterized the subsets of the parameter space in which the model is set-identifiable. For models that are not fully solvable like those studied here, one can always resort to numerical integration of the master equation in order to compute numerically the likelihood function and infer model parameters. We envision the application of the methods presented in this contribution in the design and realization of novel tests of predator-prey theory for controlled laboratory experiments~\citep{Rosenbaum2018} or field experiments~\citep{barbier2021macro,abrams2022food}. 


\section*{Acknowledgements}
We acknowledge advice and fruitful discussions with Gian Marco Palamara and the rest of members of our Unique-Priority project team. Funding was provided by grants CRISIS (PGC2018-096577-B-I00) to D.A. and J.A.C., UNIQUE (PID2021-127202NB-C21) to D.A., and PRIORITY (PID2021-127202NB-C22) to J.A.C., all funded by MCIN/AEI/10.13039/501100011033 and ``ERDF. A way of making Europe''. 

\section*{Code availability}
Code to reproduce all findings is available at 
\url{https://github.com/vankampen92/PROJECT_HOLLING_II_nD}

\appendix

\section{Steady-state distribution derivation}
\label{sec:appA00}

It can be shown that, for single resource case ($S=1$), the steady-state probability distribution of the number of free predators is a binomial $B(n_A^0,p)$ with probability parameter $p=\left(1+\frac{\theta}{\nu}\right)^{-1}$,
\begin{equation}\label{eq:binom}
P_{n_A}=\binom{n_A^0}{n_A}p^{n_A}(1-p)^{n_A^0-n_A},
\end{equation}
$n_A$ being the number of searching consumers~\citep{palamara2022implicit}. One can take advantage of knowing that distribution to rigorously derive the functional response associated to the single resource case ($S=1$), which turns out to be the Holling type II functional response~\citep{palamara2022implicit}. We here extend the binomial distribution result to a multi-resource case by solving the master equation at stationarity, which yields an analytical expression for the joint steady-state distribution $P_{\bm{n}}$. We can do this because transition probability rates~\eqref{eq:rates} are linear in population numbers~\citep{kendall1949stochastic,nisbet2004modelling}. As a result, the single-resource binomial distribution is extended to a multi-resource case, leading to a multinomial distribution for the joint probability distribution of observing a vector $(n_A,\bm{n})$ of the populations of behavioral types, as we show next.

The stationary distribution $P_{\bm{n}}$ satisfies the (infinite) system of equations defined by setting the right-hand side of Eq.~\eqref{eq:master} equal to zero. This distribution can be calculated recursively because, if the following \emph{detailed-balance conditions} hold,
\begin{equation}\label{eq:balance1}
T(\bm{n}|\bm{n}+\bm{e_i})P_{\bm{n}+\bm{e_i}} = 
T(\bm{n}+\bm{e_i}|\bm{n})P_{\bm{n}},
\end{equation}
for $i=1,\dots,S$, then all the terms in the right-hand side of Eq.~\eqref{eq:master} cancel out. This can be seen as follows. According to Eq.~\eqref{eq:rates}, detailed-balance conditions~\eqref{eq:balance1} reduce to
\begin{equation}\label{eq:balance}
\nu_i (n_i+1) P_{\bm{n}+\bm{e}_i} = 
\theta_i \bigg(n_A^0-\sum_{j=1}^S n_j\bigg) P_{\bm{n}},
\end{equation}
for $i=1,\dots,S$. Making the change $n_i\to n_i-1$ in the last equation one gets
\begin{equation*}
T(\bm{n}-\bm{e_i}|\bm{n})P_{\bm{n}} = 
T(\bm{n}|\bm{n}-\bm{e_i})P_{\bm{n}-\bm{e_i}},
\end{equation*}
which together with~\eqref{eq:balance1} cancel out the $i$-th term in the right-hand side of the master equation. Therefore, if detailed-balance conditions~\eqref{eq:balance} hold, then the probability distribution that satisfies them is the stationary distribution.

Let us apply Eq.~\eqref{eq:balance} to the $i$-th and $j$-th handling predator's populations: this yields
\begin{equation}\label{eq:recur1}
P_{\bm{n}}=\frac{\nu_i}{\theta_i}
\frac{n_i+1}{n_A^0-\sum_{k=1}^S n_k}P_{\bm{n}+\bm{e}_i},\qquad
P_{\bm{n}}=\frac{\nu_j}{\theta_j}
\frac{n_j+1}{n_A^0-\sum_{k=1}^S n_k}P_{\bm{n}+\bm{e}_j},
\end{equation}
and, subsequently,
\begin{equation*}
P_{\bm{n}+\bm{e}_i}=\frac{\nu_j}{\theta_j}\frac{\theta_i}{\nu_i}
\frac{n_j+1}{n_i+1}P_{\bm{n}+\bm{e}_j}.
\end{equation*}
Making the change $n_i\to n_i-1$ in the last equation one gets
\begin{equation}\label{eq:recur}
P_{\bm{n}}=\frac{\nu_j}{\theta_j}\frac{\theta_i}{\nu_i}
\frac{n_j+1}{n_i}P_{\bm{n}-\bm{e}_i+\bm{e}_j}.
\end{equation}
This equation is the fundamental identity we are going to use to calculate recursively the distribution $P_{\bm{n}}$, because it has the property of keeping constant the total number of handling predators $\sum_{j=1}^S n_j$ after each recursion step.

Now, apply Eq.~\eqref{eq:recur} to resources $i=1$ and $j=2$: this gives
\begin{equation*}
P_{\bm{n}}=\frac{\theta_1}{\nu_1}\frac{\nu_{2}}{\theta_{2}}
\frac{n_{2}+1}{n_1}P_{(n_1-1,n_2+1,\dots,n_S)}.
\end{equation*}
Apply again the recursion~\eqref{eq:recur} to resources $i=1$ and $j=2$ in $P_{(n_1-1,n_2+1,\dots,n_S)}$:
\begin{equation*}
P_{\bm{n}}=\left(\frac{\theta_1}{\nu_1}\frac{\nu_{2}}{\theta_{2}}\right)^2
\frac{(n_{2}+1)(n_{2}+2)}{n_1(n_1-1)}P_{(n_1-2,n_2+2,\dots,n_S)}.
\end{equation*}
By repeating the iteration $n_1-2$ additional times one gets
\begin{equation*}
P_{\bm{n}}=
\left(\frac{\theta_1}{\nu_1}\frac{\nu_{2}}{\theta_{2}}\right)^{n_1}
\binom{n_{1}+n_2}{n_1}P_{(0,n_1+n_2,\dots,n_S)},
\end{equation*}
where $\binom{m}{p}=\frac{m!}{n!(m-n)!}$ stands for the binomial coefficient $m$ choose $p$. We observe that we have written an arbitrary probability $P_{\bm{n}}$ in terms of the probability of a configuration that has no predators handling on the first resource.

The recursion can be applied to $P_{(0,n_1+n_2,\dots,n_S)}$ using the positions $i=2$ and $j=3$. After $n_{1}+n_2$ recursion steps, the number of individuals handling on the second resource will be exactly zero:
\begin{equation*}
P_{\bm{n}}=
\left(\frac{\theta_1}{\nu_1}\frac{\nu_{2}}{\theta_{2}}\right)^{n_1}
\left(\frac{\theta_{2}}{\nu_{2}}\frac{\nu_{3}}{\theta_{3}}\right)^{n_1+n_2} 
\binom{n_{1}+n_2}{n_1}
\binom{n_{1}+n_{2}+n_3}{n_{1}+n_2}P_{(0,0,n_1+n_2+n_3,\dots,n_S)},
\end{equation*}
which simplifies to
\begin{equation*}
P_{\bm{n}} =
\left(\frac{\theta_1}{\nu_1}\right)^{n_1}
\left(\frac{\theta_{2}}{\nu_{2}}\right)^{n_{2}}
\left(\frac{\nu_{3}}{\theta_{3}}\right)^{n_{1}+n_2} 
\binom{n_{1}+n_{2}+n_3}{n_{1},n_{2},n_3}
P_{(0,0,n_1+n_2+n_3,\dots,n_S)},
\end{equation*}
where we have used known relations between the binomial and the multinomial coefficients, the latter being defined as $\binom{m}{s_1,\dots,s_q}=\frac{m!}{\prod_{k=1}^q s_k!}$ (with $s_1+\dots +s_q=m$). Here, probabilities are expressed in terms of configurations in which no individual consumers feed on resources $1$ and $2$.

Iterating the process until all the populations in every class have moved to the $S$-th one, we obtain 
\begin{equation}\label{eq:interm}
P_{\bm{n}} =
\binom{\sum_{k=1}^S n_k}{n_{1},\dots,n_S}
\left(\frac{\nu_{S}}{\theta_{S}}\right)^{\sum_{k=1}^{S-1} n_k} 
\left[\prod_{k=1}^{S-1}\left(\frac{\theta_k}{\nu_k}\right)^{n_k}\right] 
P_{(0,\dots,0,\sum_{k=1}^S n_k)}.
\end{equation}
To finish our derivation we just need to evaluate $P_{(0,\dots,0,q)}$ for an arbitrary overall number $q=\sum_{k=1}^S n_k$ of predators handling on resource $S$. This can be achieved using the first identity in Eq.~\eqref{eq:recur1}, which for $i=S$ can be written, after changing $n_S\to n_S-1$, as
\begin{equation*}
P_{\bm{n}-\bm{e}_S}=
\frac{\nu_S n_S}{\theta_1(n_A^0+1-\sum_{k=1}^S n_k)}P_{\bm{n}}.
\end{equation*}
We choose $\bm{n}=(0,\dots,0,q)=q\bm{e}_S$ in the last equation. Upon $q$ iterations, we obtain
\begin{equation*}
\begin{aligned}
P_{(q-1)\bm{e}_S}&=\frac{\nu_S}{\theta_S}
\frac{q}{n_A^0+1-q}P_{q\bm{e}_S},\\
P_{(q-2)\bm{e}_S}&=\left(\frac{\nu_S}{\theta_S}\right)^2
\frac{q(q-1)}{(n_A^0+1-q)(n_A^0+2-q)}P_{q\bm{e}_S},\\
&\phantom{e}\vdots\\
P_{\bm{0}}&=\left(\frac{\nu_S}{\theta_S}\right)^{q}
\frac{q!}{(n_A^0+1-q)\cdots(n_A^0-1)n_A^0}P_{q\bm{e}_S},
\end{aligned}
\end{equation*}
where $\bm{0}=(0,\dots,0)$. Equivalently,
\begin{equation*}
P_{q\bm{e}_S}=\binom{n_A^0}{q}
\left(\frac{\theta_S}{\nu_S}\right)^{q}P_{\bm{0}}.
\end{equation*}
Setting $q=\sum_{k=1}^S n_k$ in the last equation and substituting into~\eqref{eq:interm} we obtain the joint stationary distribution $P_{\bm{n}}$ up to a normalization constant $P_{\bm{0}}$,
\begin{equation}\label{eq:interm1}
P_{\bm{n}}=
\frac{n_A^0!}{n_{1}!\cdots n_S!n_A!}
\left[\prod_{k=1}^S\left(\frac{\theta_k}{\nu_k}\right)^{n_k}\right]
P_{\bm{0}},
\end{equation}
where we have used Eq.~\eqref{eq:const} and 
\begin{equation*}
\binom{n_A^0}{\sum_{k=1}^S n_k}\binom{\sum_{k=1}^S n_k}{n_{1},\dots,n_S} =
\frac{n_A^0!}{n_{1}!\cdots n_S!n_A!}.
\end{equation*}
Observe that the distribution calculated in~\eqref{eq:interm1} up to normalization is non-zero over the hyperplane defined by $n_A+\sum_{k=1}^S n_k= n_A^0$. State vectors outside that plane have probability zero of being realized.

In the form of Eq.~\eqref{eq:interm1}, the joint stationary distribution remarkably resembles a multinomial distribution, see Eq.~\eqref{eq:multidef}. To write Eq.~\eqref{eq:interm1} as a multinomial distribution, first we have to calculate the normalization constant $P_{\bm{0}}$. It follows immediately after evaluating the probability of observing a total number $q$ of handling predators, $q=\sum_{k=1}^S n_k$, which can be done as follows.

Let $P_q$ be the probability that the overall number of handling predators, $\sum_{k=1}^S n_k$, is equal to $q$. It can be calculated (up to normalization) by summing Eq.~\eqref{eq:interm1} over the hyperplane $\sum_{k=1}^S n_k=q$:
\begin{equation*}
P_q = \sum_{\sum_i n_i=q} \frac{n_A^0!}{n_{1}!\cdots n_S!n_A!}
\left[\prod_{k=1}^S\left(\frac{\theta_k}{\nu_k}\right)^{n_k}\right]
P_{\bm{0}},
\end{equation*}
which can be written alternatively as
\begin{equation*}
P_q = \binom{n_A^0}{q}\sum_{\sum_i n_i=q} \frac{q!}{n_{1}!\cdots n_S!}
\left[\prod_{k=1}^S\left(\frac{\theta_k}{\nu_k}\right)^{n_k}\right]
P_{\bm{0}}.
\end{equation*}
According to the multinomial theorem~\citep{knuth1997art}, the sum above can be explicitly expressed as 
\begin{equation*}
\sum_{\sum_i n_i=q} \frac{q!}{n_{1}!\cdots n_S!}
\prod_{k=1}^S\left(\frac{\theta_k}{\nu_k}\right)^{n_k} = \left(\sum_{k=1}^S\frac{\theta_k}{\nu_k}\right)^{q},
\end{equation*}
which implies that
\begin{equation}\label{eq:totalhand}
P_q = \binom{n_A^0}{q}\left(\sum_{k=1}^S\frac{\theta_k}{\nu_k}\right)^{q}
P_{\bm{0}}.
\end{equation}
Therefore, if we impose normalization of the distribution $P_q$ we get
\begin{equation}\label{eq:norm}
P_{\bm{0}}^{-1} = \sum_{q=0}^{n_A^0}\binom{n_A^0}{q}
\left(\sum_{k=1}^S\frac{\theta_k}{\nu_k}\right)^{q}=
\left(1+\sum_{k=1}^S\frac{\theta_k}{\nu_k}\right)^{n_A^0},
\end{equation}
where in the last equality we have used the binomial theorem.

The joint distribution of the vector of population abundances of different behavioral types, $(n_A,n_1,\dots,n_S)$, is given by Eqs.~\eqref{eq:interm1} and~\eqref{eq:norm}. It is a multinomial distribution $\mathcal{M}(n_A^0,\{p_A,p_1,\dots,p_S\})$, as defined in Eq.~\eqref{eq:multidef}, with $n_A^0$ trials and probabilities given by Eqs.~\eqref{eq:pi} and \eqref{eq:pA}, which we reproduce here:
\begin{equation*}
p_i = \frac{\theta_i}{\nu_i}\frac{1}{1+\sum_{k=1}^S \frac{\theta_k}{\nu_k}}
\end{equation*}
for predators handling on resources $i=1,\dots,S$, and  
\begin{equation*}
p_A = \frac{1}{1+\sum_{k=1}^S \frac{\theta_k}{\nu_k}}
\end{equation*}
for free predators. Observe that $p_A+\sum_{k=1}^S p_k=1$ is satisfied. In addition, Eq.~\eqref{eq:pA} extends naturally, for more than one resource, the probability parameter $p=\left(1+\frac{\theta}{\nu}\right)^{-1}$ of the binomial distribution~\eqref{eq:binom}. 

Observe also that the average population number of each behavioral type is $\langle n_i\rangle = n_A^0 p_i$. This immediately leads to the average number of predators feeding on a focal resource of class $i$,
\begin{equation}
\langle n_i\rangle = \frac{n_A^0}{1 + \sum_{k=1}^S \frac{\theta_k}{\nu_k}}
\frac{\theta_i}{\nu_i},
\label{eq:n_i_star_HII_multi}
\end{equation}
and a mean number of free predators equal to
\begin{equation}
\langle n_A\rangle = \frac{n_A^0}{1 + \sum_{k=1}^S \frac{\theta_k}{\nu_k}}.
\label{eq:n_A_star_HII_multi}
\end{equation}
This automatically yields the multi-resource Holling type II functional response extension (see~\cite{palamara2022implicit} for the derivation using a different methodology).

In addition to the joint, steady-state probability distribution of different behavioral types, by summing over the corresponding variables we can obtain the marginal distributions of each single type, which are binomials $B(n_A^0,p_i)$ \citep{loeve1977elementary}. In particular, from our derivation above we find that the distribution of the total number of handling consumers is a binomial $B(n_A^0,1-p_A)$: using Eqs.~\eqref{eq:norm} and~\eqref{eq:pA}, the probability that $\sum_{k=1}^S n_k=q$, which is given by Eq.~\eqref{eq:totalhand}, can be rewritten as
\begin{equation*}
P_q=\binom{n_A^0}{q}(1-p_A)^q p_A^{n_A^0-q}.
\end{equation*}
This immediately implies that the distribution of the number of free predators, $n_A=n_A^0-\sum_{k=1}^S n_k = n_A^0-q$, follows a binomial distribution $B(n_A^0,p_A)$,
\begin{equation*}
P_{n_A}=\binom{n_A^0}{n_A}p_A^{n_A} (1-p_A)^{n_A^0-n_A},
\end{equation*}
as expected~\citep{palamara2022implicit}.

If we were to use the multinomial steady-state distribution~\eqref{eq:multidef} with probabilities given by Eqs.~\eqref{eq:pi} and~\eqref{eq:pA}, to estimate feeding rates vía ML (as above, we assume that $n_A^0$ is fixed and not subject to MLE), we could write the following exhaustive summary: $\bm{\kappa}=\left(\theta_1/\nu_1,\dots,\theta_S/\nu_S\right)$, which is tantamount to $\bm{\kappa}=\left(\alpha_1/\nu_1,\dots,\alpha_S/\nu_S\right)$ because resource densities $x_{R_i}$ are kept constant (chemostatic conditions). Since the log-likelihood is determined by $S$ ratios and there are $2S$ individual rates ($\alpha_1,\dots,\alpha_S,\nu_1,\dots,\nu_S$) to be estimated, the model is structurally unidentifiable (parameter redundant) if MLE is conducted using the steady-state distribution.

\section{Derivation of the PDE satisfied by the generating function}
\label{sec:appA0}

This Appendix is devoted to derive the PDE~\eqref{eq:PDE} satisfied by the generating function~\eqref{eq:Gdef} in the out-of-equilibrium multi-resource Holling type II case. First we multiply Eq.~\eqref{eq:master} by $z_1^{n_1}\cdots z_S^{n_S}$ and sum over all non-negative integer vectors $\bm{n}$. The sum can be split as 
\begin{equation*}
\frac{\partial G}{\partial t}=A+B,
\end{equation*}
with terms $A$ and $B$ defined as
\begin{equation*}
A := \sum_{n_1,\dots,n_S\ge 0}\sum_{i=1}^S z_1^{n_1}\cdots z_S^{n_S}
\left[T(\bm{n}|\bm{n}+\bm{e_i})P_{\bm{n}+\bm{e_i}}
-T(\bm{n}-\bm{e_i}|\bm{n})P_{\bm{n}}\right]
\end{equation*}
and
\begin{equation*}
B := \sum_{n_1,\dots,n_S\ge 0}\sum_{i=1}^S z_1^{n_1}\cdots z_S^{n_S}
\left[T(\bm{n}|\bm{n}-\bm{e_i})P_{\bm{n}-\bm{e_i}}
-T(\bm{n}+\bm{e_i}|\bm{n})P_{\bm{n}}\right],
\end{equation*}
respectively. 

First, focus on the $A$ term and the transition probabilities given by Eq.~\eqref{eq:rates}, which are linear functions of the population numbers. In order to evaluate the sums in terms of $G$ and its partial derivatives, we can use the identity
\begin{equation*}
\frac{\partial G}{\partial z_i} = \sum_{n_1,\dots,n_S\ge 0} (n_i+1)
P_{\bm{n}+\bm{e}_i} z_1^{n_1}\cdots z_S^{n_S},
\end{equation*}
which follows from the definition~\eqref{eq:Gdef} and the fact that $P_{-\bm{e}_i}(t)=0$ for every index $i$, which has to be imposed for population numbers to remain positive. We also take into account the identity
\begin{equation}\label{eq:id1}
z_i\frac{\partial G}{\partial z_i} = \sum_{n_1,\dots,n_S\ge 0} n_i
P_{\bm{n}} z_1^{n_1}\cdots z_S^{n_S},
\end{equation}
which directly follows from the definition of the generating function. Using both identities, we can immediately write the term $A$ as function of $z_i$ and the first order partial derivative of the generating function $G$ with respect to $z$, as follows:
\begin{equation*}
A = \sum_{n_1,\dots,n_S\ge 0} \sum_{i=1}^S [\nu_i(n_i+1)P_{\bm{n}+\bm{e}_i}
-\nu_i n_i P_{\bm{n}}] 
 z_1^{n_1}\cdots z_S^{n_S} = 
\sum_{i=1}^S \nu_i(1-z_i)\frac{\partial G}{\partial z_i}. 
\end{equation*}

Second, and taking into account the expressions of transition probabilities given by Eqs.~\eqref{eq:rates}, we observe that the $B$ term can be written as the following sum:
\begin{equation*}
B = \sum_{n_1,\dots,n_S\ge 0}\sum_{i=1}^S z_1^{n_1}\cdots z_S^{n_S}\biggl\{
n_A^0\theta_i\left(P_{\bm{n}-\bm{e_i}}-P_{\bm{n}}\right)
+\theta_i\biggl[P_{\bm{n}}\sum_j n_j-\biggl(n_i-1+\sum_{j\ne i}n_j\biggr)
P_{\bm{n}-\bm{e}_i}\biggr]\biggr\}.
\end{equation*}
The following identities also follow from the definition of the generating function:
\begin{equation*}
\begin{aligned}
z_i^2\frac{\partial G}{\partial z_i} &= \sum_{n_1,\dots,n_S\ge 0} (n_i-1)
P_{\bm{n}-\bm{e}_i} z_1^{n_1}\cdots z_S^{n_S},\\
z_i z_j\frac{\partial G}{\partial z_j} &= \sum_{n_1,\dots,n_S\ge 0} n_j
P_{\bm{n}-\bm{e}_i} z_1^{n_1}\cdots z_S^{n_S}\,\,\,(i\ne j).
\end{aligned}
\end{equation*}
They are based also in the property $P_{-\bm{e}_i}(t)=0$ for all $i$, which must hold for all $i$ so that configurations with negative populations are assigned probability zero. Using these two identities and~\eqref{eq:id1}, together with
\begin{equation*}
z_i G = \sum_{n_1,\dots,n_S\ge 0} P_{\bm{n}-\bm{e}_i} z_1^{n_1}\cdots z_S^{n_S},
\end{equation*}
which follows also from $P_{-\bm{e}_i}(t)=0$, we find an expression for the $B$ term in terms of $G$ and its derivatives,
\begin{equation*}
B = \sum_{i=1}^S \theta_i(1-z_i) \biggl(-n_A^0 G
+\sum_{j=1}^S z_j\frac{\partial G}{\partial z_j}\biggr).
\end{equation*}
Gathering the contributions of $A$ and $B$ terms, we obtain the PDE~\eqref{eq:PDE} satisfied by the generating function $G(\bm{z},t)$.

\section{ML estimation with equal sampling times}
\label{sec:appB}

In this section we show that sampling data at equal times, $t_1=\dots=t_T=t$ ($t>0$), makes the model unidentifiable. Here we write the multinomial distribution as
\begin{equation*}
P_{\bm{n}}(n_A^0,\{p_i\}) = C(\bm{n})\left(1-\sum_{i=1}^S p_i\right)^{n_A}
\prod_{i=1}^S p_i^{n_i},
\end{equation*}
where $\bm{n}=(n_1,\dots,n_S)$ is the vector of handling predators abundances, $C(\bm{n})$ is the multinomial coefficient, $n_A=n_A^0-\sum_{j=1}^Sn_j$ is the number of free predators, and probability parameters $p_j$ are given, for $j=1,\dots,S$, by Eq.~\eqref{eq:pj}. Assume equal rates, $\alpha_j=\alpha$ and $\nu_j=\nu$ for $j=1,\dots,S$. Then the multinomial probability parameters are expressed as
\begin{equation*}
p_j(t) = \frac{x_{R_j}\alpha}{\nu} \cdot \frac{1-e^{-\left(\nu+\alpha \Lambda_R\right)t}}
{1+\alpha \Lambda_R/\nu}.
\end{equation*}
Here we have defined the total resource density $\Lambda_R:=\sum_{i=1}^S x_{R_i}$. Then we can write $p_i=x_{R_i}h(t;\alpha,\tau)$, where
\begin{equation*}
h(t;\alpha,\tau)=\alpha \tau (1-e^{-t/\tau})
\end{equation*}
and $\tau=(\nu+\alpha \Lambda_R)^{-1}$ is the characteristic time scale.

The log-likelihood function is expressed as
\begin{equation*}
\log\mathcal{L}(\alpha,\nu\vert M(t)) = T\log C(\bm{n}) + \sum_{k=1}^T
\left\{n_{AR,k}\log h(t;\alpha,\tau) + n_{A,k}\log(1-\Lambda_R h(t;\alpha,\tau))
+ \sum_{i=1}^S n_{i,k}\log x_{R_i} \right\},
\end{equation*}
where we have used that $p_i=x_{R_i}h(t;\alpha,\tau)$ and defined the $k$-th sampled total handing predators abundance as the sum $n_{AR,k}=\sum_{i=1}^S n_{i,k}$, and the $k$-th sampled number of free predators as $n_{A,k}=n_A^0-\sum_{i=1}^S n_{i,k}=n_A^0-n_{AR,k}$ ---the last equality holds for every sample, $k=1,\dots,T$. Observe that the dependence on model parameters $\alpha$ and $\nu$ has been isolated in function $h$. Therefore, the first and last terms in the log-likelihood function do not contribute to derivatives with respect to those parameters. 

Thus, the critical points of the log-likelihood satisfy the equations 
\begin{equation*}
\begin{aligned}
\frac{\partial}{\partial \alpha}\log \mathcal{L} =
\frac{\partial h}{\partial\alpha}\sum_{k=1}^T
\left(\frac{n_{AR,k}}{h}-\frac{n_{A,k} \Lambda_R}{1-\Lambda_R h}\right) &=0,\\
\frac{\partial}{\partial \nu}\log \mathcal{L} =
\frac{\partial h}{\partial\nu}\sum_{k=1}^T
\left(\frac{n_{AR,k}}{h}-\frac{n_{A,k} \Lambda_R}{1-\Lambda_R h}\right)&=0.
\end{aligned}
\end{equation*}
Then one can check that the system formed by $\frac{\partial h}{\partial\alpha}=0$ and $\frac{\partial h}{\partial\nu}=0$ has no solution satisfying at the same time the obvious restrictions $\alpha,\nu,t>0$. Therefore the two conditions for critical points reduces to one single equation,
\begin{equation*}
\sum_{k=1}^T\left(\frac{n_{AR,k}}{h}-\frac{n_{A,k} \Lambda_R}{1-\Lambda_R h}\right)=0,
\end{equation*}
from which we can eliminate the function $h(t;\alpha,\tau)$ to obtain the condition to be satisfied by model parameters at the critical points. This defines a manifold in the parameter space, given by

\vspace*{-1mm}
\begin{equation}\label{eq:mani}
\frac{\alpha\left[1-e^{-\left(\nu+\alpha \Lambda_R\right)t}\right]}{\nu+\alpha \Lambda_R}=
\frac{\langle n_{AR}\rangle_T}{\Lambda_R n_A^0},
\end{equation}
where we have defined $\langle n_{AR}\rangle_T :=\frac{1}{T}\sum_{k=1}^T n_{AR,k}$ as the
average across samples of handling consumers' total abundance. The quantity $\langle n_{AR}\rangle_T$, relative to the product $\Lambda_R n_A^0$, determines only one single restriction to be satisfied by model parameters, in the $(\alpha,\nu)$ space. This immediately implies that the two parameters cannot be uniquely determined, yielding (for every $t>0$) an infinite number of critical points satisfying Eq.~\eqref{eq:mani}, as we wanted to prove. In the case of exactly equal sampling times (either out of equilibrium or at stationarity), the model remains set-identifiable. At least two sampling times must be distinct in order for $\alpha$ and $\nu$ being non-redundant.

Finally, observe that, in the limit $t\to\infty$, the manifold defined by Eq.~\eqref{eq:mani} reduces to the straight line whose intercept is equal to zero and whose slope is
\begin{equation*}
\frac{\nu}{\alpha}=\frac{\langle n_A\rangle_T}{\langle n_{AR}\rangle_T}\Lambda_R,
\end{equation*}
as confirmed by the numerical optimization of the equilibrium log-likelihood function ---here $\langle n_{A}\rangle_T :=\frac{1}{T}\sum_{k=1}^T n_{A,k}=n_A^0-\langle n_{AR}\rangle_T$. The slope of the line of infinite maxima is determined by the average number of free predators across samples, relative to the average total number of handling predators, times the overall density of resources. This is consistent with Eq.~\eqref{eq:slope} if, at equilibrium, $\langle n_A\rangle$ and $\langle n_{AR}\rangle$ are estimated from data by their corresponding averages across samples, i.e., by $\langle n_{A}\rangle_T$ and $\langle n_{AR}\rangle_T$, respectively.

\section{Macroscopic model dynamics}
\label{sec:appC}

In this section we integrate the macroscopic dynamics 
$$
\frac{dx_i}{dt} = -\nu x_i +\theta_i\Bigg ( n_A^0 - \sum_{j=1}^S x_i \Bigg),
$$
for $x_i=\langle n_i\rangle$ the mean-field abundance of consumers feeding on resource $i=1,\dots,S$. Once the abundances are solved as function of $t$, one can obtain the average number of free consumers as $x_A=\langle n_A\rangle = n_A^0 -\sum_{i=1}^S x_i$. Initially, all handling predator abundances are set equal to zero ($x_i(0)=0)$, which means that all consumers are initially free, $x_A(0)=n_A^0$. 

This model is linear and can be written in a compact form as
$$
\frac{d\bm{x}}{dt} = n_A^0 \bm{\theta} - J \bm{x},
$$
for vector $\bm{\theta}=(\theta_i)=\alpha(x_{R_i})$, matrix $J=\nu I+\bm{\theta}\bm{1}^T$ (which, as a rank-one perturbation of a diagonal matrix, is invertible by the Sherman-Morrison's formula) and initial condition $\bm{x}(0)=\bm{0}$. The equilibrium abundances are given by $\bm{x}^{\star}=n_A^0 J^{-1}\bm{\theta}$. If we write $\bm{y}=\bm{x}-\bm{x}^{\star}$ then the dynamics reduces to $\frac{d\bm{y}}{dt} = - J \bm{y}$ with initial condition $\bm{y}(0)=-\bm{x}^{\star}$. In that form, it can be solved by matrix exponentiation,
$$
\bm{y}(t)=e^{-J t}\bm{y}(0)=-n_A^0 e^{-J t} J^{-1}\bm{\theta}.
$$
Thus, in terms of the original variables $\bm{x}$, we obtain
\begin{equation}
\bm{x}(t)=\bm{y}(t)+\bm{x}^{\star}=n_A^0(I-e^{-J t})J^{-1}\bm{\theta}.
\label{eq:xt}
\end{equation}
The problem is reduced to calculate the exponential $e^{-Jt}$ and the inverse $J^{-1}$.

In order to sum the exponential series, we write $J=\nu H$ and compute the powers of $H=I+\bm{h}\bm{1}^T$ with $\bm{h}=\nu^{-1}\bm{\theta}$. Given that $H$ is a rank-one perturbation of the identity, we expect that
$$
H^n = I + c_n \bm{h}\bm{1}^T,
$$
for a sequence $c_n$ to be determined, which obviously satisfies $c_1=1$. Now compute $H^{n+1}$ as
$$
H^{n+1}=(I + c_n \bm{h}\bm{1}^T)(I + \bm{h}\bm{1}^T) =
I + ( (\bm{1}^T\bm{h} + 1)c_n + 1 )\bm{h}\bm{1}^T,
$$
from which we obtain the recurrence $c_{n+1}=(\bm{1}^T\bm{h}+1)c_n + 1$ to be solved with initial condition $c_1=1$. Using that $\bm{1}^T\bm{h}=\nu^{-1}\bm{1}^T\bm{\theta}=Q/\nu$, and solving the recurrence for coefficients $c_n$, we obtain
$$
J^n = \nu^n I + \frac{(\nu+Q)^n-\nu^n}{Q}\bm{\theta}\bm{1}^T.
$$
Then, by summing the series, it is straightforward to write the exponential of $-tJ$ as, again, a rank one perturbation of a diagonal matrix:
\begin{equation}
e^{-J t} = e^{-\nu t}I + \frac{1}{Q}\left(e^{-(\nu+Q)t}-e^{-\nu t}\right)\bm{\theta}\bm{1}^T.
\label{eq:expJt}
\end{equation}

On the other hand, by the Sherman-Morrison's formula, 
$$
J^{-1}=\nu^{-1}(I-(\nu+Q)^{-1}\bm{\theta}\bm{1}^T),
$$
and therefore $J^{-1}\bm{\theta}=\frac{1}{\nu+Q}\bm{\theta}$. Substituting this result and Eq.~\eqref{eq:expJt} into Eq.~\eqref{eq:xt} we finally obtain
\begin{equation}
\bm{x}(t)=\frac{n_A^0}{\nu+Q}(1-e^{(\nu+Q)t})\bm{\theta}.
\label{eq:xsol}
\end{equation}
The mean abundance of free predators is $x_A(t) = n_A^0 - \bm{1}^T\bm{x}(t)$:
\begin{equation}
x_A(t)=\frac{n_A^0}{\nu+Q}(\nu+Qe^{(\nu+Q)t})=
n_A^0(\tau \nu + (1-\tau\nu)e^{-t/\tau}),
\end{equation}
for $\tau=(\nu+Q)^{-1}$. Observe that, asymptotically, $\lim_{t\to\infty}\bm{x}(t)=\frac{n_A^0}{\nu+Q}\bm{\theta}$ and $\langle n_A\rangle_{\infty}=\lim_{t\to\infty}x_A(t)=\frac{n_A^0\nu}{\nu+Q}$. The total abundance of handling predators at equilibrium, $\langle n_{AR}\rangle_{\infty}=\lim_{t\to\infty}\bm{1}^T\bm{x}(t)=\frac{n_A^0 Q}{\nu+Q}$. Naturally, because $Q=\alpha\sum_{i=1}^S x_{R_i}$, the expressions obtained for the averages $\langle n_A\rangle_{\infty}$ and $\langle n_{AR}\rangle_{\infty}$ satisfy Eq.~\eqref{eq:slope}. Thus, the ratio $\nu/\alpha$ can be estimated using steady-state averages. Additional information is necessary to fully determine individual rates $\alpha$ and $\nu$.

\renewcommand\thefigure{D.\arabic{figure}} 
\setcounter{figure}{0}

\begin{figure*}[t!]
\centering
\includegraphics[width=0.6\textwidth]{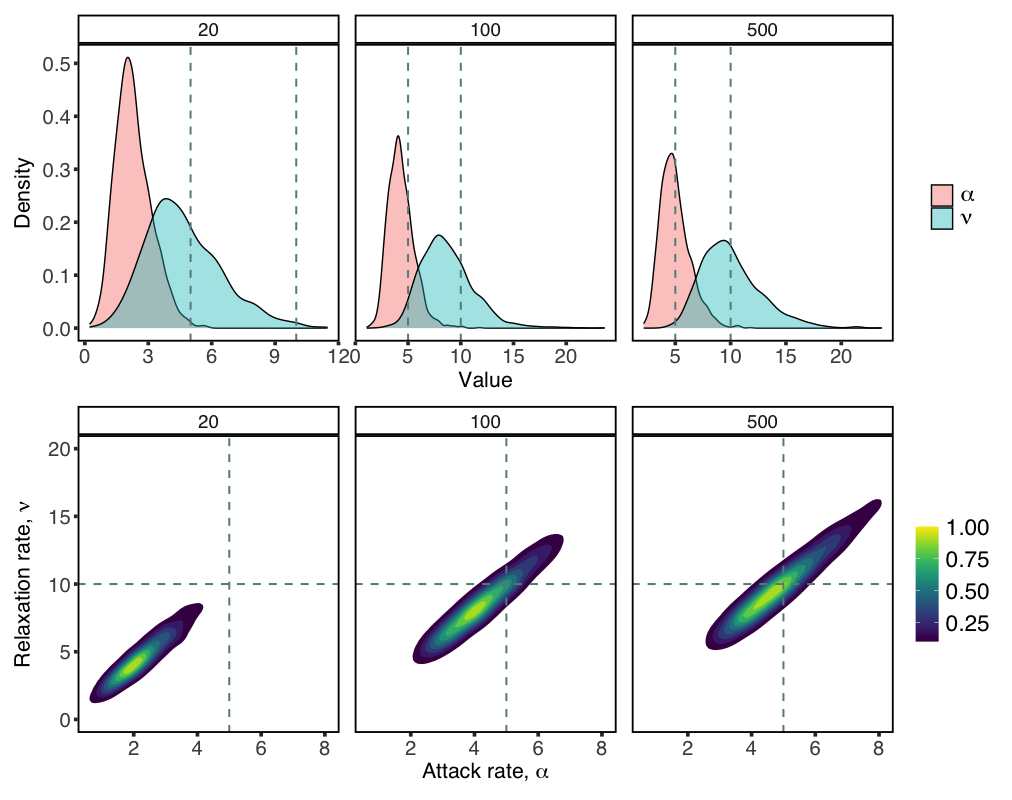}
\caption{\label{fig:appD} \textbf{Distributions of parameter estimates based on the deterministic dynamics, for different number of initial consumers}. Same as Fig.~\ref{fig:dist} for a estimation procedure based on the deterministic model. We generated 1500 stochastic realizations from $t=0$ to $t=100\tau$. Then the slope at $t=0$ in each stochastic realization, which equals $Q=\alpha\sum_{i=1}^S x_{R_i}$, was used to estimate $\alpha$. For that purpose, we fit a linear model to the function $n_A(t)/n_A^0$ using the first 20 times in each realization. Second, we used the asymptotic values $\langle n_A\rangle_{\infty}$ and $\langle n_{AR}\rangle_{\infty}$ to estimate the ratio $\nu/\alpha$: from each realization we average $n_A(t)$ and $n_{AR}(t)$ neglecting the first 500 temporal points, enough to reach stationarity. Compared to distributions in Fig.~\ref{fig:dist}, we obtain with this method wider distributions and more biased estimates, especially for small $n_A^0$. Parameters are the same as in Fig.~\ref{fig:eq}.} 
\end{figure*}

A possible solution would be having external information about the characteristic time of the dynamics, $\tau$. Here we propose using the initial decay of the number of free predators as function of $t$. For $t\ll \tau$, we can approximate $x_A(t)=n_A^0(1-Qt)$, so we can use the slope of $x_A(t)$ at $t=0$ to determine $\alpha$ via $Q=\alpha\sum_{i=1}^S x_{R_i}$. We have done so using stochastic realizations. and we estimated $\alpha$ using the slope at $t=0$ of each realization. We generated long-time realizations to estimate the ratio $\nu/\alpha$ from Eq.~\eqref{eq:slope}, i.e., $\nu/\alpha=(\langle n_A\rangle_{\infty}/\langle n_{AR}\rangle_{\infty})\sum_{i=1}^R x_{R_i}$. Results are reported in Fig.~\ref{fig:appD}. Observe than estimates obtained from this approach are more biased than the estimates obtained by maximum likelihood (compare with Fig.~\ref{fig:dist}), yielding very poor estimates for $n_A^0$ small. In addition, the width of the parameter distributions does not reduce as $n_A^0$ increases. This is not surprising because ML estimators are, under natural regularity conditions, those with minimal variance. More importantly, the estimation procedure cannot be based solely on steady-state abundances: additional information is needed from an out-of-equilibrium regime (in this case, the rate at which the abundance of free predators decays for shorter times).



\bibliographystyle{plainnat}
\bibliography{references}
\end{document}